\pdfoutput=1

\documentclass[letterpaper,twocolumn,10pt]{article}
\usepackage{usenix2019_v3,epsfig}
\usepackage[bf,small]{caption} 

\usepackage[normalem]{ulem}

\usepackage{url}
\usepackage{graphicx}
\usepackage{listings}
\usepackage{epstopdf}
\usepackage{color}
\usepackage{subfigure}
\usepackage{xspace}
\usepackage{latexsym}
\makeatletter
\newif\if@restonecol
\makeatother

\usepackage[ruled,vlined,linesnumbered]{algorithm2e}
\usepackage{algorithmic}
\usepackage{enumerate}

\usepackage{makecell}

\usepackage{fontawesome}

\usepackage{wrapfig}
\usepackage{setspace}
\usepackage{blindtext}
\usepackage{amsfonts}
\usepackage{multirow}
\usepackage{pifont}
\usepackage{mathrsfs}
\usepackage{mathtools}
\usepackage{relsize}
\usepackage{ulem}
\usepackage{comment}
\usepackage{footnote}
\usepackage{tikz}
\usetikzlibrary{calc}
\makesavenoteenv{tabular}
\makesavenoteenv{table}

\usepackage{enumitem}

\newif{\ifSubmit}
\newif{\ifFinal}
\newif{\ifDraft}
\Drafttrue

\ifSubmit
\newcommand{\alicomment}[1]{}
\newcommand{\yuecomment}[1]{}
\newcommand{\vcomment}[1]{}
\else
\newcommand{\alicomment}[1]{\noindent\textcolor{magenta}{\bf Ali: #1}}

\newcommand{\aocomment}[1]{\noindent\textcolor{RedOrange}{Ao: #1}}

\newcommand{\vcomment}[1]{\textcolor{green}{\textbf{Vasily: #1}}}
\fi

\newcommand{\added}[1]{\noindent\textcolor{RubineRed}{#1}}

\newcommand{\proj}{\textsc{FaaSNet}}
\newcommand{\cloud}{Alibaba Cloud}
\newcommand{\fc}{Function Compute}

\newcommand{\base}{\textit{baseline}}
\newcommand{\opt}{\textit{on-demand}}
\newcommand{\optt}{\textit{DADI+P2P}}
\newcommand{\kraken}{\textit{Kraken}}
\newcommand{\fastlsmd}{I/O efficient format}

\usepackage{booktabs}
\usepackage{siunitx}

\usepackage{zref}
\usepackage{cleveref}
\crefformat{section}{\S#2#1#3}

\DeclarePairedDelimiter\floor{\lfloor}{\rfloor}
\microtypecontext{spacing=nonfrench}

\usepackage{balance}

\usepackage[numbers]{natbib}
\begin{document}

\title{\Large {\proj}: Scalable and Fast Provisioning of Custom Serverless Container Runtimes at Alibaba Cloud Function Compute\thanks{This is the preprint version of a paper published in USENIX ATC 2021.}}

\author{
	{\rm Ao Wang$^{1}$, Shuai Chang$^{2}$, Huangshi Tian$^{3}$, Hongqi Wang$^{2}$, Haoran Yang$^{2}$, Huiba Li$^{2}$, }\\
	{\rm  Rui Du$^{2}$, Yue Cheng$^1$}\\
	{{{$^{1}$}George Mason University} ~~~{$^{2}$}{Alibaba Group} ~~~{$^{3}$}{Hong Kong University of Science and Technology}}\\
}
\maketitle

\pagestyle{plain}

\begin{abstract}

Serverless computing, or Function-as-a-Service (FaaS), 
enables a new way of building and scaling applications by allowing users to deploy fine-grained functions while providing fully-managed resource provisioning and auto-scaling. 
Custom FaaS container support is gaining traction as it enables better control over OSes, versioning, and tooling for modernizing FaaS applications.
However, providing rapid container provisioning introduces non-trivial challenges for FaaS providers, since container provisioning is costly, and real-world FaaS workloads exhibit highly dynamic patterns.

In this paper, we design {\proj}, a highly-scalable middleware system for accelerating FaaS container provisioning.
{\proj} is driven by the workload and infrastructure requirements of the FaaS platform at one of the world's largest cloud providers, {\cloud} {\fc}. 
{\proj} enables scalable container provisioning via a lightweight, adaptive function tree (FT) structure. 
{\proj} uses an I/O efficient, on-demand
fetching mechanism to further reduce provisioning costs at scale.
We implement and integrate {\proj} in {\cloud} {\fc}.
Evaluation results show that {\proj}: (1) finishes provisioning $2,500$ function containers on $1,000$ virtual machines in $8.3$ seconds, 
(2) scales $13.4\times$ and $16.3\times$ faster than {\cloud}'s 
current FaaS platform and a state-of-the-art P2P container registry (Kraken), respectively,
and (3) sustains
a bursty workload using $75.2\%$ less time than an optimized baseline.

\end{abstract}
\vspace{-10pt}
\section{Introduction}
\label{sec:intro}
\vspace{-6pt}

In recent years, a new cloud computing model called serverless computing or Function-as-a-Service (FaaS)~\cite{berkeley_serverless} has emerged. Serverless computing enables a new way of building and scaling applications and services by allowing developers to break traditionally monolithic server-based applications into finer-grained cloud functions. Developers write function logic while the service provider performs the notoriously tedious tasks of provisioning, scaling, and managing the backend servers~\cite{gray_stop} that the functions run on. 


Serverless computing solutions are growing in popularity and finding their way into both commercial clouds (e.g., AWS Lambda~\cite{lambda}, Azure Functions~\cite{azure_func}, Google Cloud Functions~\cite{google_func} and Alibaba Cloud Function Compute\footnote{We call {\fc} throughout the paper.}~\cite{alibaba_fc}, etc.) and open source projects (e.g., OpenWhisk~\cite{openwhisk}, Knative~\cite{knative}). While serverless platforms such as AWS Lambda and Google Cloud Functions support functions packaged as {\small\texttt{.zip}} archives~\cite{lambda_package}, this deployment method 
poses constraints for FaaS applications with a lack of flexibility. 
One constraint is a maximum  package size limit (of up to 250 MB uncompressed for AWS Lambda functions). 

A recent trend is the support of packaging and deploying cloud functions using custom container images~\cite{aws_lambda_container,google_cloud_run,fc_custom_container,openfaas}. This approach is desirable as it greatly enhances usability, portability, and tooling support:
(1)~Allowing cloud functions to be deployed as custom container runtimes enables many interesting application scenarios~\cite{custom_container_doc}, which heavily rely on large dependencies such as machine learning~\cite{tensorflow_osdi16, ray_osdi18}, data analytics~\cite{spark_nsdi12, wukong_socc20, wukong_pdsw19}, and video processing~\cite{excamera_nsdi17, sprocket_socc18}; this would not have been possible with limited function package sizes.
(2)~Container tooling (e.g., Docker~\cite{docker}) simplifies the software development and testing procedures; therefore, developers who are familiar with container tools can easily build and deploy FaaS applications using the same approach. 
(3)~This approach will enable new DevOps features such as incremental update (similar to rolling update in Kubernetes~\cite{rolling_update_k8s}) for FaaS application development.




A potential benefit that makes the FaaS model appealing is the fundamental resource elasticity---ideally, 
a FaaS platform must allow a user application to scale up to tens of thousands of cloud functions on demand, in seconds, with no advance notice. 
However, providing rapid container provisioning for custom-container-based FaaS infrastructure introduces non-trivial challenges.

First, FaaS workloads exhibit highly dynamic, bursty patterns~\cite{serverless_in_the_wild}. To verify this, we analyzed a FaaS workload from a production serverless computing platform managed by one of the world's largest cloud providers, \emph{{\cloud}}. 
We observe that
a single application's function request throughput (in terms of concurrent invocation requests per second or RPS) can spike up to more than a thousand RPS with a peak-to-trough ratio of more than $500\times$ (\cref{subsub:burstiness}). 
A FaaS platform typically launches many virtualized environments---in our case at {\fc}, virtual machines (VMs) that host and isolate containerized functions---on demand to serve request surges~\cite{lambda, firecracker_nsdi20, peeking_atc18}.
The bursty workload will create a network bandwidth bottleneck when hundreds of VMs that host the cloud functions are pulling the same container images from the backing store (a container registry or an object store). As a result, the high cost of the container startup process\footnote{A container startup process typically includes downloading the container image manifest and layer data, extracting layers, and starting the container runtime; in our paper we call the  startup process \emph{container provisioning}.} makes it extremely difficult for FaaS providers to deliver the promise of high elasticity. 

Second, custom container images are large in sizes. For example, more than $10\%$ of the containers in Docker Hub are larger than 1.3~GB~\cite{docker_hub_analysis}. Pulling large container images from the backing store would incur significant cold startup latency, which can be up to several minutes (\cref{subsub:container_pull}) if the backing store is under high contention. 
Existing solutions 
cannot be directly applied to our FaaS platform.
Solutions such as Kraken~\cite{uber_kraken}, DADI~\cite{DADI}, and Dragonfly~\cite{dragonfly} use peer-to-peer (P2P) approaches to accelerate container provisioning at scale; however, they require one or multiple dedicated, powerful servers serving as root nodes for data seeding, metadata management, and coordination.
Directly applying these P2P-based approaches to our existing FaaS infrastructure is not an ideal solution due to the following reasons.
(1) It would require extra, dedicated, centralized components, thus increasing the total cost of ownership (TCO) for the provider while introducing a performance bottleneck. 
(2) Our FaaS infrastructure uses a dynamic pool of resource-constrained VMs to host containerized cloud functions for strong isolation; a host VM
may join and leave the pool at any time. This dynamicity requires a highly adaptive solution, which existing solutions fail to support.

To address these challenges, we present {\proj}, a lightweight and adaptive  middleware system for accelerating serverless container provisioning.
{\proj} enables high scalability by decentralizing the container provisioning process across host VMs 
that are organized in 
function-based tree structures. 
A \emph{function tree} (\emph{FT}) is a logical, tree-based network overlay. A FT consists of multiple host VMs and allows provisioning 
of container runtimes or code packages
to be decentralized across all VMs in a scalable manner.
{\proj} enables high adaptivity via a tree balancing algorithm that dynamically adapts the FT topology in order to accommodate VM joining and leaving.

Note that the design of {\proj} is driven by the specific workload and infrastructure requirements of {\cloud} {\fc}. For example, {\fc} uses containers inside VMs to provide strong tenant-level isolation. A typical FaaS VM pool has thousands of small VM instances. The scale of the FaaS VM pool and the unique characteristics of FaaS workloads determine that: 
(1)~a centralized container storage would not scale;
and (2)~existing container distribution techniques may not work well in our  environment as they have different assumptions on both workload types and underlying cluster resource configurations.

We make the following contributions in this paper.
\vspace{-6pt}
\begin{itemize}[noitemsep,leftmargin=*]
    \item We present the design of a FaaS-optimized, custom container provisioning middleware system called {\proj}. At {\proj}'s core is an adaptive function tree abstraction that avoids central bottlenecks.
    \item We implement and integrate {\proj} in 
    {\cloud} {\fc}. {\proj} is, to the best of our knowledge, the first FaaS container provisioning system from a cloud provider with published technical details.
    \item We deploy {\proj} in {\cloud} {\fc} and evaluate {\proj} extensively using both production workloads and microbenchmarks. Experimental results show that {\proj}: finishes provisioning $2,500$ function containers within 8.3 seconds (only $1.6\times$ longer than that of provisioning a single container), scales $13.4\times$ and $16.3\times$ faster than {\cloud}'s current FaaS platform and a state-of-the-art P2P registry (Kraken), respectively, and sustains a bursty workload using $75.2\%$ less time than an optimized baseline.
    \item We release {\proj}'s FT and an anonymized dataset containing production FaaS cold start traces.
\end{itemize}
\vspace{-4pt}

\vspace{-12pt}
\section{Background and Motivation}
\label{sec:moti}
\vspace{-6pt}

In this section, we first provide an overview of the FaaS container workflows in {\cloud} {\fc}.
We then present a motivational study on the FaaS workloads to highlight the bursty patterns and their demands of a scalable and elastic FaaS container runtime provisioning system.

\vspace{-12pt}
\subsection{FaaS Container Workflows in {\cloud} {\fc}}
\label{subsec:abcloud}
\vspace{-6pt}


\begin{figure}[t]
\begin{center}
\includegraphics[width=.35\textwidth]{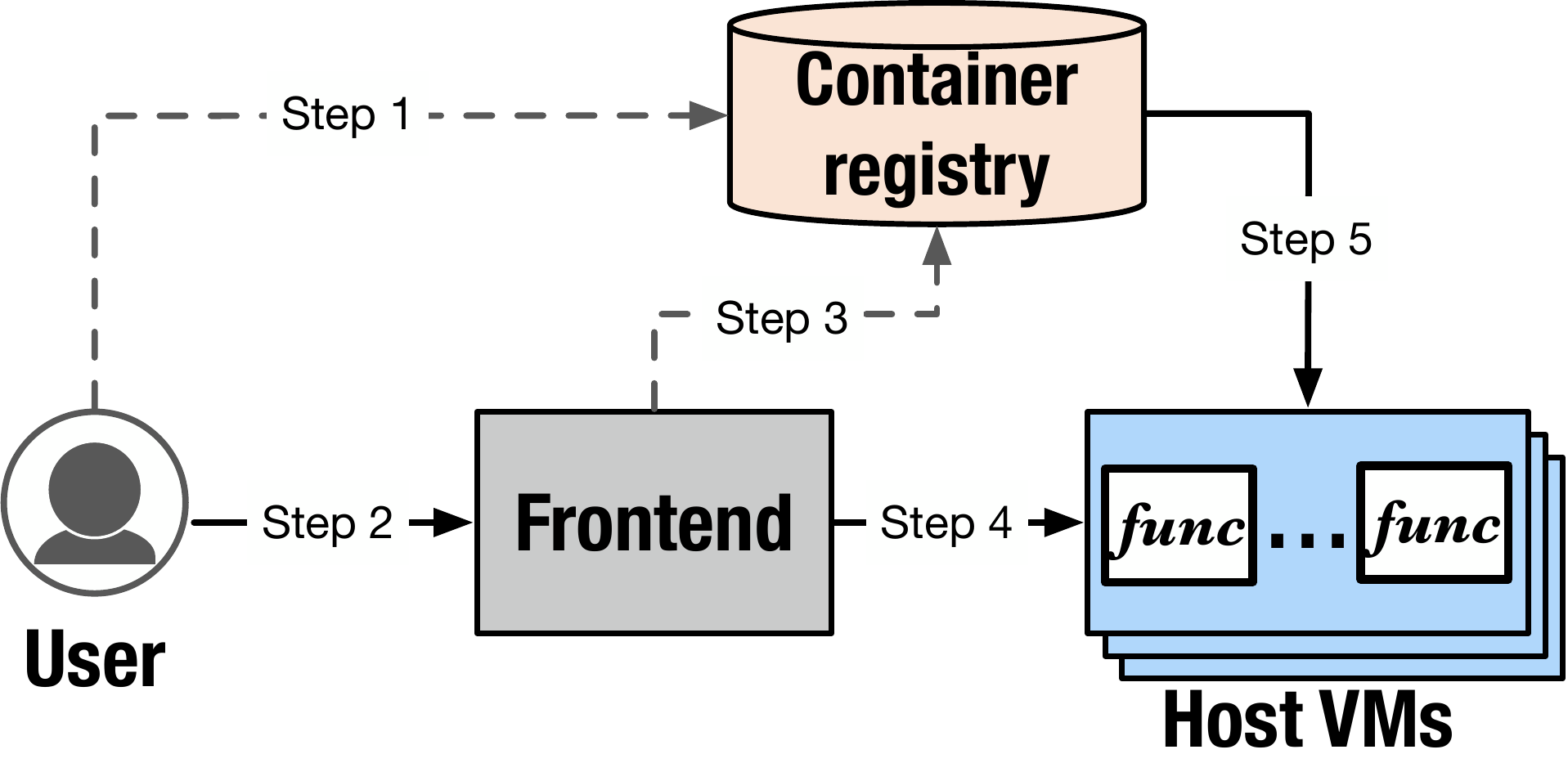}
\vspace{-10pt}
\caption{Overview of {\cloud}'s FaaS container workflows. 
}
\label{fig:background}
\end{center}
\vspace{-25pt}
\end{figure}

\begin{figure*}
\begin{center}
\includegraphics[width=1\textwidth]{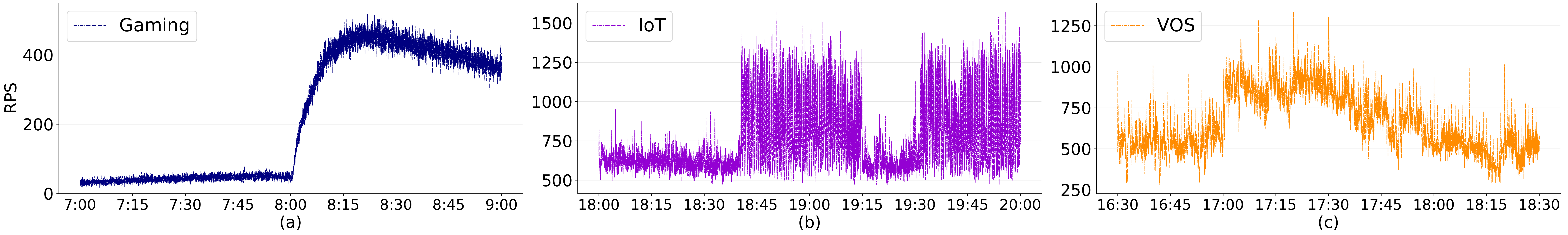}
\vspace{-15pt}
\caption{2-hour throughput timelines of example FaaS applications. 
}
\label{fig:thpt_trace}
\end{center}
\vspace{-15pt}
\end{figure*}

\begin{figure*}[t]
\begin{center}
\subfigure[Image pull latency.] {
\includegraphics[width=.24\textwidth]{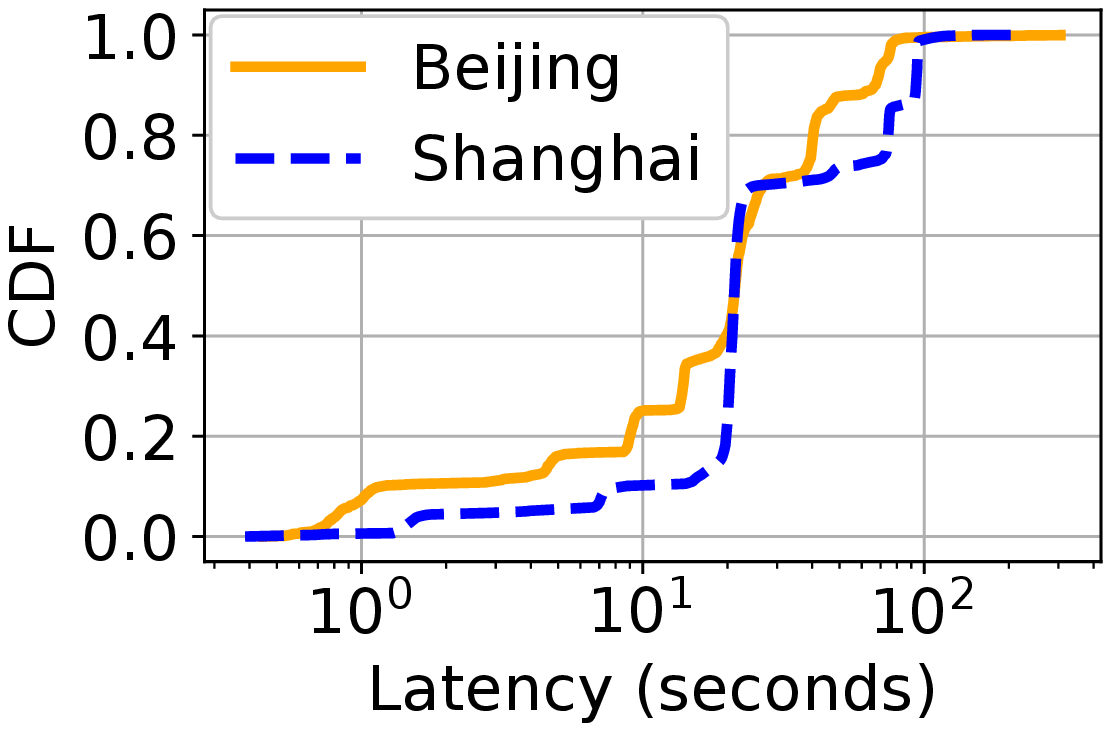}
\label{fig:pull_latency}
}
\hspace{-9pt}
\subfigure[
Proportion of image pull in cold start.
] {
\includegraphics[width=.24\textwidth]{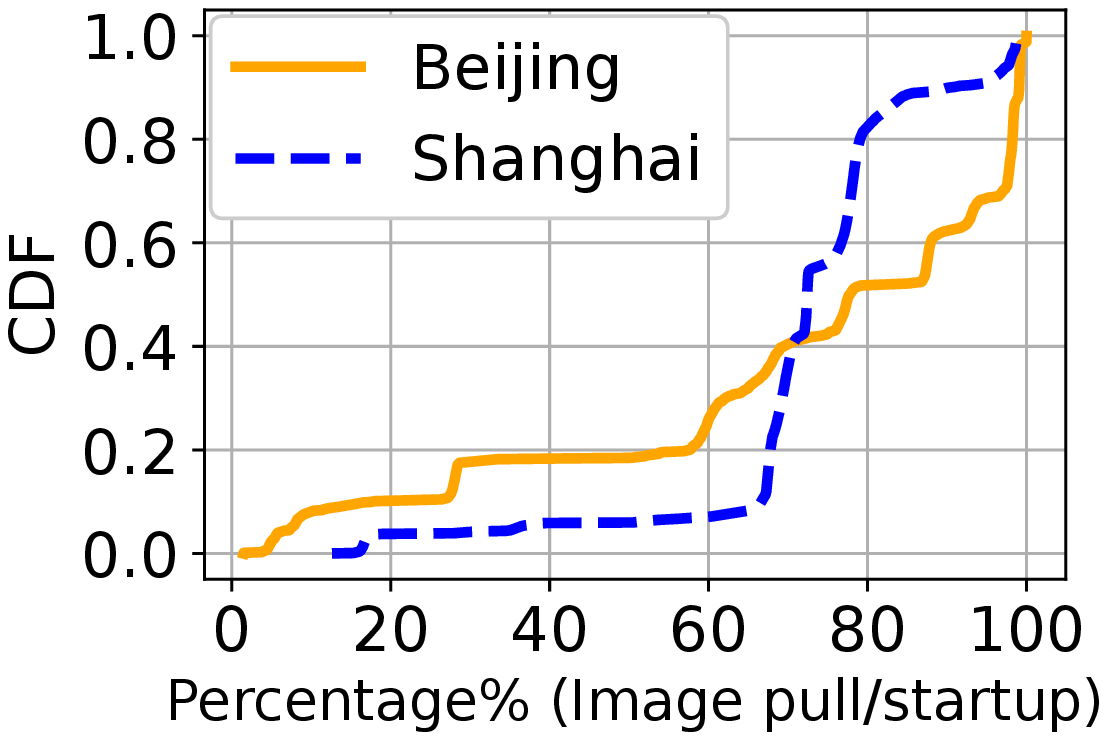}
\label{fig:pull_percentile}
}
\hspace{-8pt}
\subfigure[Inter-arrival time of cold starts.] {
\includegraphics[width=.24\textwidth]{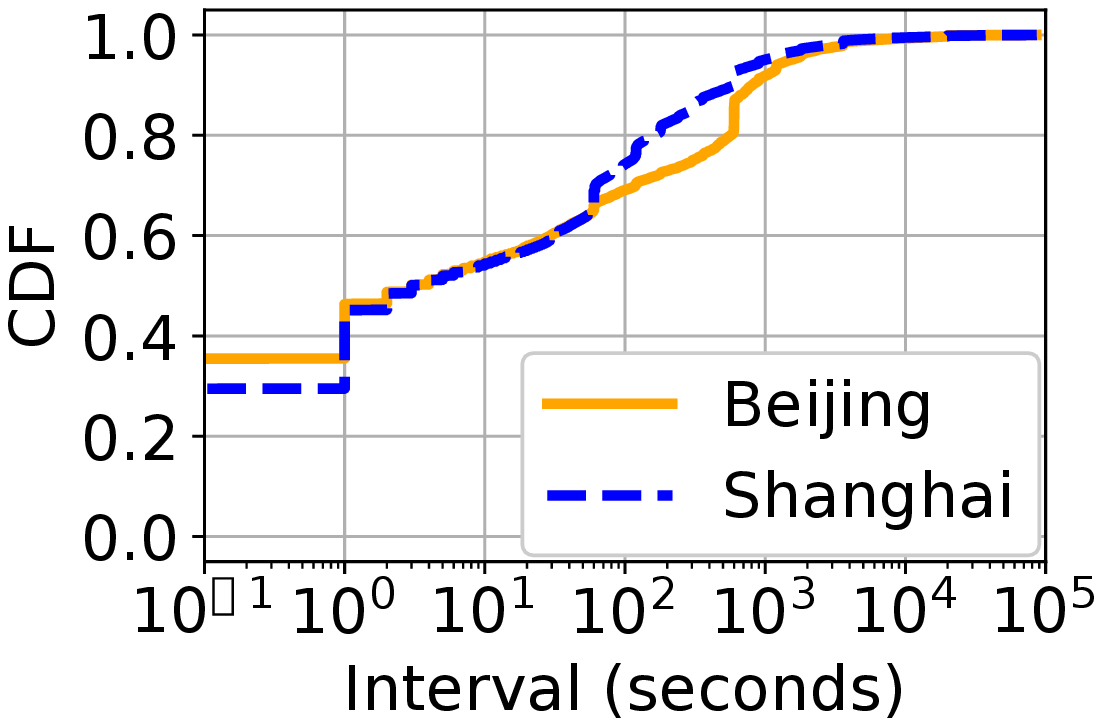}
\label{fig:function_arrival_rate}
}
\hspace{-9pt}
\subfigure[Function duration.] {
\includegraphics[width=.24\textwidth]{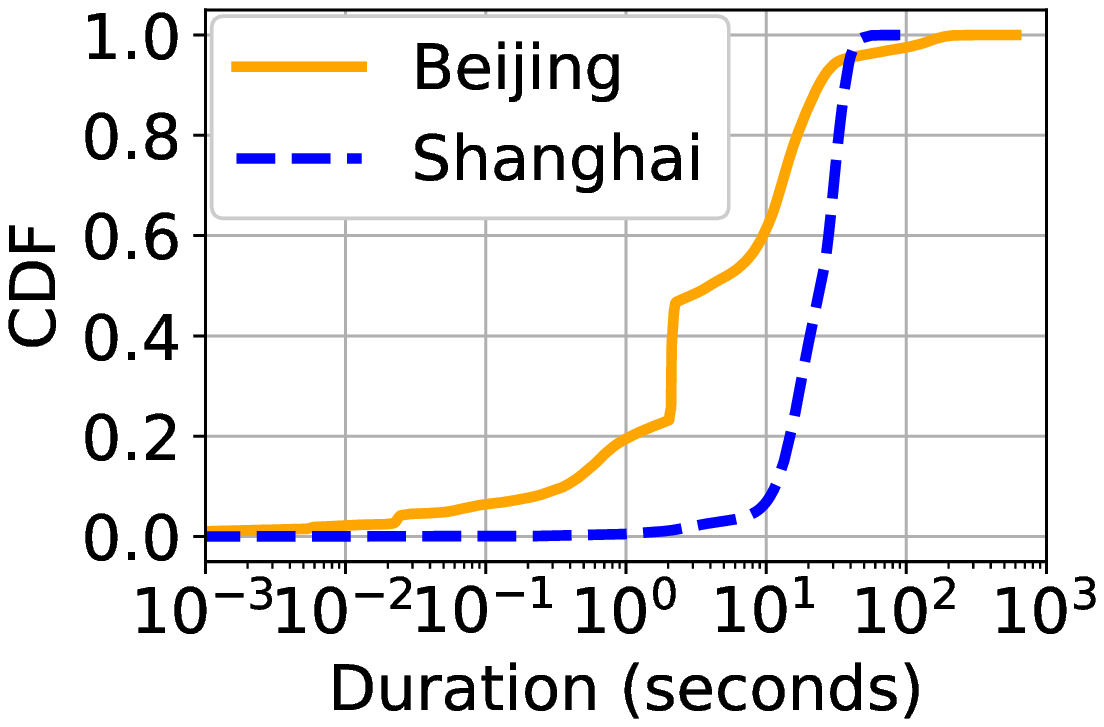}
\label{fig:function_duration}
}
\vspace{-10pt}
\caption{Performance characteristics of container image pulls (a, b) and function invocations (c, d) in CDF.}
\label{fig:cluster_pull}
\end{center}
\vspace{-15pt}
\end{figure*}

{\fc} allows users to build and deploy FaaS applications using custom container images and container tools. Figure~\ref{fig:background} shows a typical workflow of function deployment and invocation. 
To deploy (or update) a containerized function, a user sends a {\small\texttt{create}}/{\small\texttt{update}} request  in order to push the container image to a centralized container registry (Step 1 in Figure~\ref{fig:background}). 
To invoke a deployed function, the user sends {\small\texttt{invoke}} requests to the frontend gateway (Step 2), which checks the user's request and the status of the container image in the registry (Step 3). 
The frontend then forwards the requests to the backend FaaS VM cluster for servicing the requests (Step 4).
Finally, host VMs create function containers and pull their container data from the registry (Step 5). 
After all the previous steps are successfully completed, host VMs become ready and start serving the invocation requests.

\vspace{-10pt}
\subsection{Workload Analysis}
\label{subsec:workload}
\vspace{-6pt}

Step 5 in Figure~\ref{fig:background}, container runtime provisioning, must be fast and scalable in order to enable high elasticity for FaaS workloads. 
To obtain a better understanding of the workload requirements, we analyze the workload traces that were collected from {\fc}.

\vspace{-10pt}
\subsubsection{Workload Burstiness}
\label{subsub:burstiness}
\vspace{-6pt}

FaaS providers charge users using a fine-grained, pay-per-use pricing model---they bill on a per invocation basis (e.g., \$0.02 per 1 million invocations for AWS Lambda) and charge the CPU and memory bundle resource usage at the millisecond level. This property is attractive to a broad class of applications that exhibit highly fluctuating and sometimes unpredictable loads; compared to traditional VM-based deployment approach that charges even when the VM resources are idle, FaaS is more cost-effective as tenants do not pay when the load is zero. 
Therefore, we analyzed the workload traces 
and verified that bursty behaviors are indeed common. Figure~\ref{fig:thpt_trace} reports the behaviors of three representative FaaS applications: gaming, IoT, and VOS (video processing).  

Figure~\ref{fig:thpt_trace}(a) shows that a request spike shoots from 22 to 485 RPS
with a peak-to-trough ratio of $22\times$.
As well as being bursty, IoT and VOS show different patterns. 
As shown in Figure~\ref{fig:thpt_trace}(b), IoT exhibit a sustained throughput of around 682 RPS, but the throughput suddenly increased to more than 1460 RPS; the peak throughput lasts for about 40 minutes and the second peak starts 15 minutes after the first peak ends. 
Whereas for VOS (Figure~\ref{fig:thpt_trace}(c)), for the first 30 minutes, it observes an average throughput of 580 RPS with a maximum (minimum) throughput of 982 (380) RPS; the average throughput increases to 920 RPS at 30 minutes, and gradually reduces back to an average of 560 RPS. 

\vspace{-12pt}
\paragraph{Implication 1:}
\emph{Such dynamic behaviors require scalable and resilient provisioning of large numbers of function containers to rapidly smooth out the latency spikes that a FaaS application may experience during a request burst.}

\vspace{-10pt}
\subsubsection{Cold Start Costs of Containerized Functions}
\label{subsub:container_pull}
\vspace{-6pt}

Next, we focus on cold start costs of containerized functions. A cold start, in our context, refers to the first-ever invocation of a custom-container-based function; a cold start latency is typically long, ranging from a few seconds to a few minutes 
as it requires the FaaS provider to fetch the image data and start the container runtime before executing the function. 
As noted in prior work~\cite{serverless_in_the_wild, sock_atc18, sand_atc18, peeking_atc18}, the high cold start penalty 
is a notorious roadblock to FaaS providers as it hurts elasticity.
The cold start issue is exacerbated when custom container feature with sizeable dependencies is supported.

We analyzed the container downloading costs in two 
FaaS regions, \textit{Beijing} and \textit{Shanghai}, managed by {\fc}. We retrieved a 15-day log, which recorded the statistics of the function container registry and reported the performance characteristics of 712,295 cold start operations for containerized functions.
As shown in Figure~\ref{fig:cluster_pull}(a), for \textit{Beijing}, about $57\%$ of the image pulls see a latency longer than 45 seconds, while for \textit{Shanghai} more than $86\%$ of the image pulls take at least 80 seconds.

We next examined the proportion of time spent on image pull with respect to the total function cold start latency.
Figure~\ref{fig:cluster_pull}(b) shows that more than $50\%$ and $60\%$ of function invocation requests spend at least $80\%$ and $72\%$ of the overall function startup time on pulling container images, for \textit{Beijing} and \textit{Shanghai} respectively. 
This indicates that the cost of image pull dominates most functions' cold start costs. 


To put the cold start costs into perspective, we further inspected cold starts' inter-arrival time and function duration.
Figure~\ref{fig:function_arrival_rate} plots the interval distribution of consecutive cold start requests. In both of the two regions, about $49\%$ of function cold starts have an inter-arrival time less than 1 second, implying a high frequency of cold start requests. 
As shown in Figure~\ref{fig:function_duration}, about $80\%$ of the function executions in \textit{Beijing} region are longer than 1 second; in \textit{Shanghai} region, about $80\%$ of the function duration is less than $32.5$ seconds, with a $90^{th}$ percentile of $36.6$ seconds and a $99^{th}$ percentile of $45.6$ seconds. This distribution indicates that cold start costs are of the same magnitude as the function duration, stressing a need for optimizing container startups. 

\vspace{-12pt}
\paragraph{Implication 2:}
\emph{Optimizing the performance of container provisioning will provide a huge benefit on reducing the cold start costs of container-based cloud functions. }
\vspace{-10pt}
\section{{\proj} Design}
\label{sec:design}
\vspace{-4pt}

\vspace{-4pt}
\subsection{Design Overview}
\label{subsec:overview}
\vspace{-6pt}

\begin{figure}
\begin{center}
\includegraphics[width=.48\textwidth]{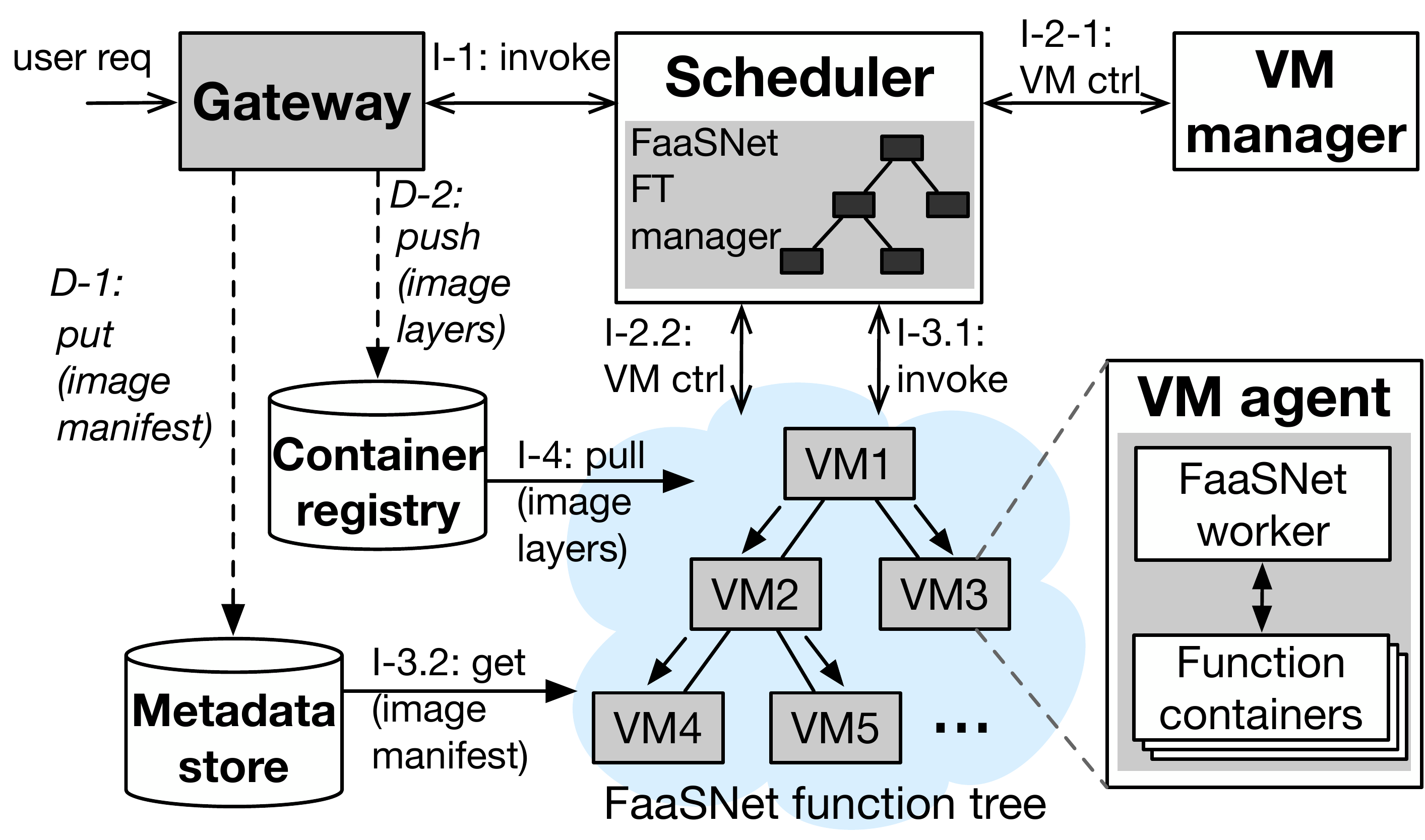}
\vspace{-12pt}
\caption{{\proj} architecture. Our work is in the gray boxes. 
Function invocation requests (solid arrow: invoke, VM ctrl, get(image manifest), and pull(image layers)) are online operations.  
{\proj} minimizes the operation of I-4: pull(image layers) and efficiently decentralizes container provisioning (i.e., image load and container start) across VMs.
Function deployment requests (dashed arrow, italic font: {\textit{D-1: put(image manifest)}} and {\textit{D-2: push(image layers)}}) are offline operations.
}
\label{fig:arch}
\end{center}
\vspace{-20pt}
\end{figure}

This section provides a high-level overview of {\proj}'s architecture. 
Figure~\ref{fig:arch} illustrates the architecture of the FaaS platform running {\proj}. 
{\proj} decentralizes and parallelizes container provisioning\footnote{While this paper mainly focuses on container runtime provisioning, {\proj} supports  provisioning of both containers and code packages.} 
across VMs. {\proj} introduces an abstraction called \emph{function trees} (\emph{FTs}) to enable efficient container provisioning at scale.
{\proj} integrates a \emph{FT manager} component and a \emph{worker} component into our existing FaaS scheduler and VM agent for coordinated FT management. Next, we describe the main components in our FaaS platform.

A \emph{gateway} is responsible for (1) tenant identity access management (IAM) authentication, (2) forwarding the function invocation requests to the FaaS scheduler, and (3) converting regular container images to the I/O efficient data format.

A \emph{scheduler} is responsible for serving function invocation requests. We integrate a {\proj} \emph{FT manager} into the scheduler to manage \emph{function trees} (\cref{subsec:ft}), or \emph{FTs} for short, through FT's {\small\texttt{insert}} and {\small\texttt{delete}} APIs. 
A FT is a binary tree overlay that connects multiple host VMs to form a fast and scalable container provisioning 
network. Each VM runs a FaaS VM agent, which is responsible for VM-local function management. We integrate a {\proj} \emph{worker} into the VM agent for container provisioning tasks. 

On the function invocation path, the scheduler first communicates with a \emph{VM manager} to scale out the the active VM pool from a free VM pool, if there are not enough VMs or all VMs that hold an instance of the requested function are busy.
The scheduler then queries its local FT metadata and sends RPC requests to  {\proj} workers of the FT to start the container provisioning process (\cref{subsec:integration}). 
The container runtime provisioning process is effectively decentralized and parallelized across all VMs in a FT that do not yet have a container runtime locally provisioned.
The scheduler sits off the critical path while {\proj} workers
fetch function container layers on demand and creates the container runtime (\cref{subsec:on_demand}) from the assigned peer VMs in parallel. 

As described in \cref{subsec:abcloud}, on the function deployment path,
the gateway converts a function's regular container image into an \emph{I/O efficient} format (\cref{subsec:on_demand}) by pulling the regular image from a tenant-facing container registry, compresses the image layers block-by-block, creates a metadata file (an image manifest) that contains the format-related information,
and writes the converted layers and its associated manifest to an {\cloud}-internal container registry and a metadata store, respectively. 

\vspace{-10pt}
\subsection{Function Trees}
\label{subsec:ft}
\vspace{-6pt}


We make the following design choices when designing FTs.
(1)~A function has a separate FT; that is, {\proj} manages FTs at function granularity.
(2)~FTs have decoupled data plane and control plane; that is, each VM worker in a FT has equivalent, simple role of container provisioning (data plane), and the global tree management (control plane) to the scheduler (\cref{subsec:integration}).
(3)~{\proj} adopts a balanced binary tree structure that can dynamically adapt to workloads.

These design choices are well aligned with  {\cloud}'s existing FaaS infrastructure and are attuned to achieve three goals: 
(1)~minimizes the I/O load of container image and layer data downloading on backing container registry, 
(2)~eliminates the tree management bottleneck and data seeding bottleneck of a central root node,
and (3)~adapts when VMs join and leave dynamically.

\begin{wrapfigure}{htbp}{0.38\columnwidth}
\vspace{-12pt}
\centering
\includegraphics[width=0.38\columnwidth]{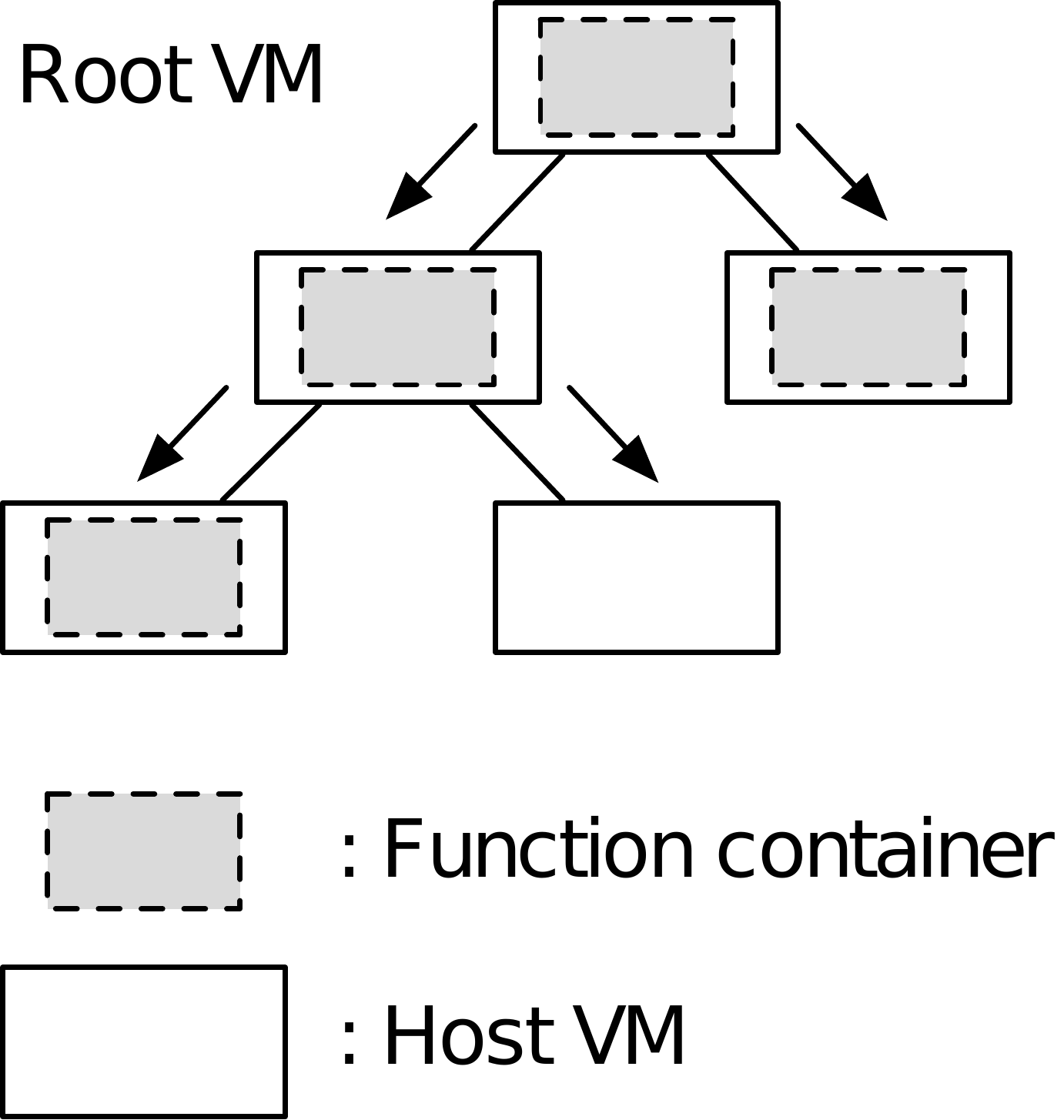}
\vspace{-15pt}
\caption{
An example {\proj} FT.
}
\vspace{-10pt}
\label{fig:ft}
\end{wrapfigure}

\noindent\textbf{Managing Trees at Function Granularity.}
{\proj} manages a separate, unique tree for each function that has been invoked at least once and has not been reclaimed. 
Figure~\ref{fig:ft} illustrates the topology of a three-level FT that spans five host VMs. 
Function container images are streamed from the root VM of the tree downwards until reaching the leaf nodes.

\vspace{-12pt}
\paragraph{Balanced Binary Trees.}
At {\proj}'s core 
is a balanced binary tree. In a binary tree, except for the root node and leaf nodes, each tree node (in our case a host VM)\footnote{We use ``node''/``VM'' interchangeably when describing tree operations.} has one incoming edge and two outgoing edges.
This design can effectively limit the number of concurrent downloading operations per VM to avoid a network contention. 
A balanced binary tree with $N$ nodes has a height of $\floor*{log(N)}$. This is desirable as a balanced binary tree guarantees that the image and layer data of a function container would traverse at most $\floor*{log(N)}$ hops from the top to the bottom. This 
%
is critical as the height of a FT would affect the efficiency of data propagation.
Furthermore, the structure of a balanced binary tree can dynamically change in order to accommodate the dynamicity of the workloads. 
To this end, {\proj} organizes each FT as a balanced binary tree. 
The \emph{FT manager} (Figure~\ref{fig:arch}) calls two APIs, {\small\texttt{insert}} and {\small\texttt{delete}}, to dynamically grow or shrink a FT.

\vspace{-12pt}
\subparagraph{\texttt{insert:}}
The very first node of a FT is inserted as a root node. The FT manager tracks the number of child nodes that each tree node has via BFS (breadth-first search) and stores all nodes that has 0 or 1 child in a queue. To insert a new node, the FT manager picks the first node from the queue as the parent of the new node. 

\begin{figure}
\begin{center}
\includegraphics[width=.45\textwidth]{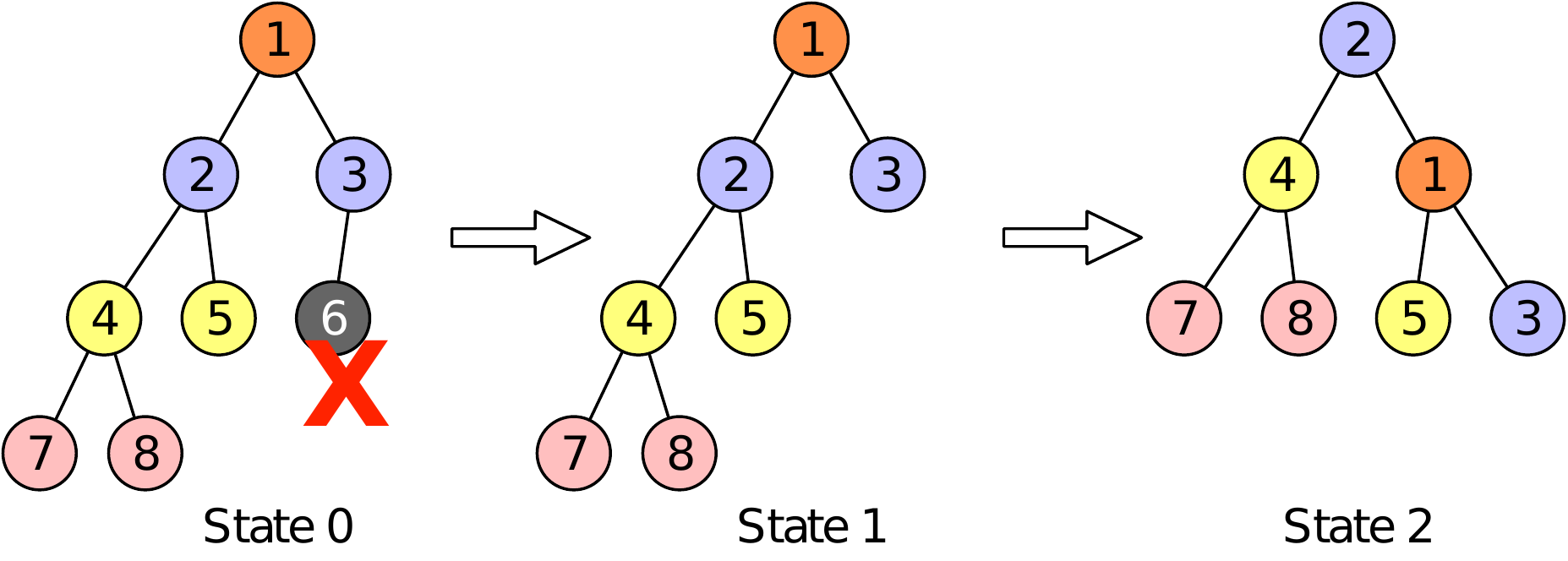}
\vspace{-5pt}
\caption{
An example \emph{right\_rotate} operation. 
The FT manager detects that Node 6 was reclaimed and calls {\small\texttt{delete}} to remove it. Removal of Node 6 causes an imbalance, which triggers a \emph{right\_rotate} rebalancing operation. 
The FT manager then performs right rotation by marking Node 2 as the new root and marking Node 5 as Node 1's left subtree.
}
\label{fig:right_rotate}
\end{center}
\vspace{-20pt}
\end{figure}

\begin{figure}
\begin{center}
\includegraphics[width=.45\textwidth]{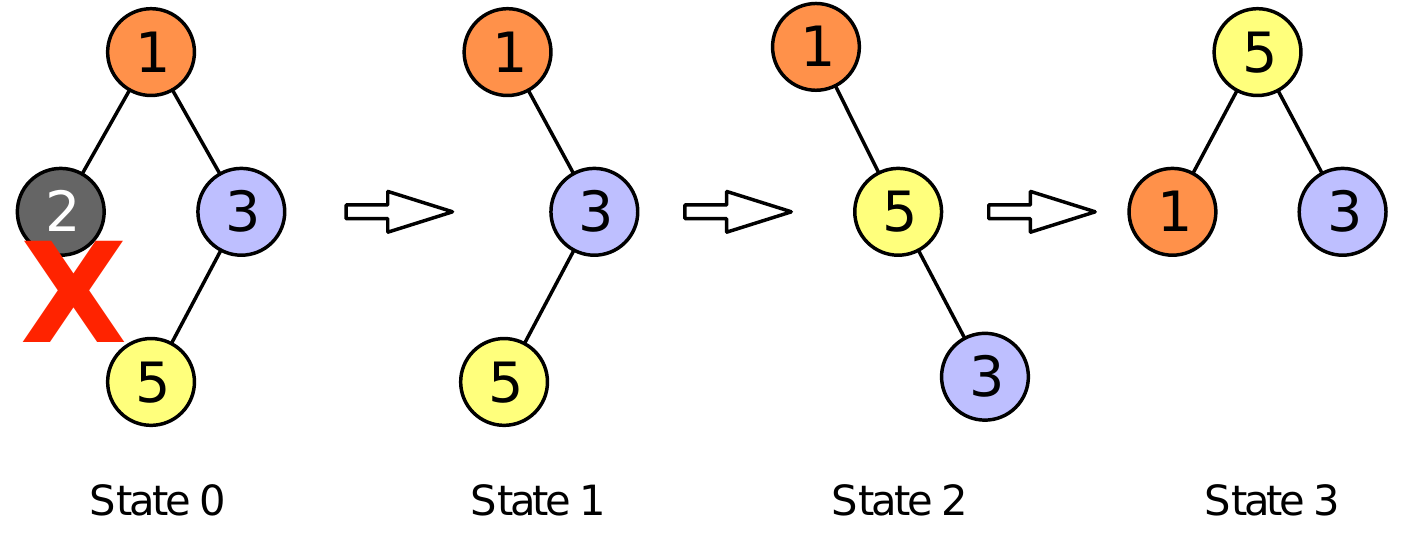}
\vspace{-5pt}
\caption{
An example \emph{right\_left\_rotate} operation. 
FT manager detects that Node 2 gets reclaimed and calls {\small\texttt{delete}} to remove it. Removal of Node 2 triggers a rebalancing operation. FT manager first right-rotates the right subtree of Node 1 by marking Node 5 as the parent of Node 3. FT manager then performs a \emph{left\_rotate} by marking Node 5 as the root.
}
\label{fig:right_left_rotate}
\end{center}
\vspace{-20pt}
\end{figure}

\vspace{-12pt}
\subparagraph{\texttt{delete:}} 
The scheduler may reclaim a VM that has been idling for a period of time (15-minute in {\cloud} configuration). Thus, FaaS VMs have a limited lifespan. To accommodate VM leaving caused by reclamation, the FT manager calls {\small\texttt{delete}} to delete a reclaimed VM. The {\small\texttt{delete}} operation rebalances the structure of FT if needed. Different from a binary search tree such as an AVL-tree or a red-black tree, nodes in a FT do not have a comparable key (and its associated value).
Therefore, our tree-balancing algorithm only needs to hold one invariant---a balancing operation is triggered only if the height difference between any node's left and right subtree is larger than 1. 
The FT implements four methods to handle all imbalance situations---\emph{left\_rotate}, \emph{right\_rotate}, \emph{left\_right\_rotate}, and \emph{right\_left\_rotate}. Due to the space limit, we omit the details of the tree balancing algorithms. 
Figure~\ref{fig:right_rotate} and Figure~\ref{fig:right_left_rotate} show the process of \emph{right\_rotate} and \emph{right\_left\_rotate} operations, respectively.

\vspace{-10pt}
\subsection{Function Tree Integration}
\label{subsec:integration}
\vspace{-6pt}

In this section, we describe how we integrate the FT scheme into {\cloud}'s FaaS platform. The integration spans two components of our existing FaaS platform, the scheduler and the VM agent. Specifically, we integrate {\proj}'s FT manager into {\cloud}'s FaaS scheduler and {\proj}'s VM worker into {\cloud}'s FaaS VM agent, respectively (Figure~\ref{fig:arch}).
The scheduler manages VMs of a FT via the FT manager. 
The scheduler starts a {\proj} worker on each VM agent.
A {\proj} worker is responsible for (1) serving scheduler's commands to perform tasks of image downloading and container provisioning, and (2) managing the VM's function containers. 

\vspace{-12pt}
\paragraph{FT Metadata Management.}
The scheduler maintains an in-memory mapping table that records the \textit{<$function_{ID}$, FT>} key-value pairs, which map a function ID to its associated FT data structure.
A FT data structure manages a set of in-memory objects representing functions and VMs to keep track of information such as a VM's \textit{address:port}.
The scheduler is sharded and is highly available. Each scheduler shard periodically synchronizes its in-memory metadata state with a distributed metadata server that runs etcd~\cite{etcd}.

\begin{figure}[t]
\begin{center}
\includegraphics[width=.48\textwidth]{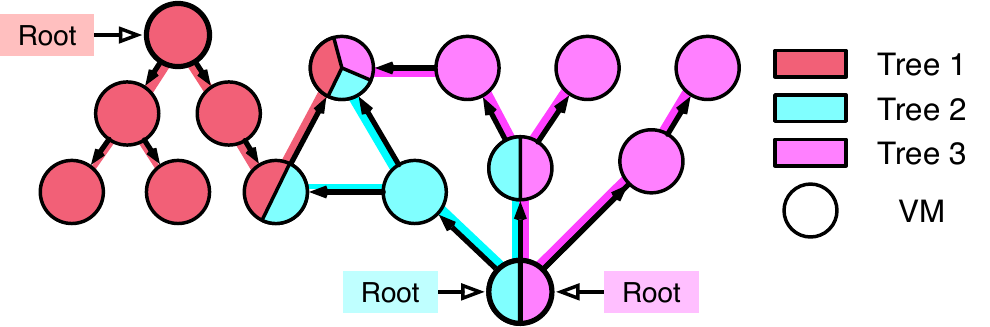}
\vspace{-10pt}
\caption{
Example function placement on VMs. 
The color codings of trees and tree edges are red for tree 1 (left), blue for tree 2 (center), and purple for tree 3 (right). Arrows denote provisioning flows.
}
\label{fig:tree-node-overlap}
\end{center}
\vspace{-20pt}
\end{figure}

\vspace{-12pt}
\paragraph{Function Placement on VMs.}
For efficiency, {\proj} allows one VM to hold multiple functions that belong to the same user. 
{\fc} uses a binpacking heuristic that assigns as many functions as possible in one VM host as long as the VM has enough memory to host the functions.
As such, a VM may be involved in the topologies of multiple overlapping FTs.
Figure~\ref{fig:tree-node-overlap} shows an example of a possible FT placement. In order to avoid network bottlenecks, {\proj} limit the number of functions that can be placed on a VM---in our deployment we set this limit to 20. We discuss a proposal of the FT-aware placement in \cref{sec:discussion}.

\begin{figure}[t!]
\begin{center}
\includegraphics[width=.48\textwidth]{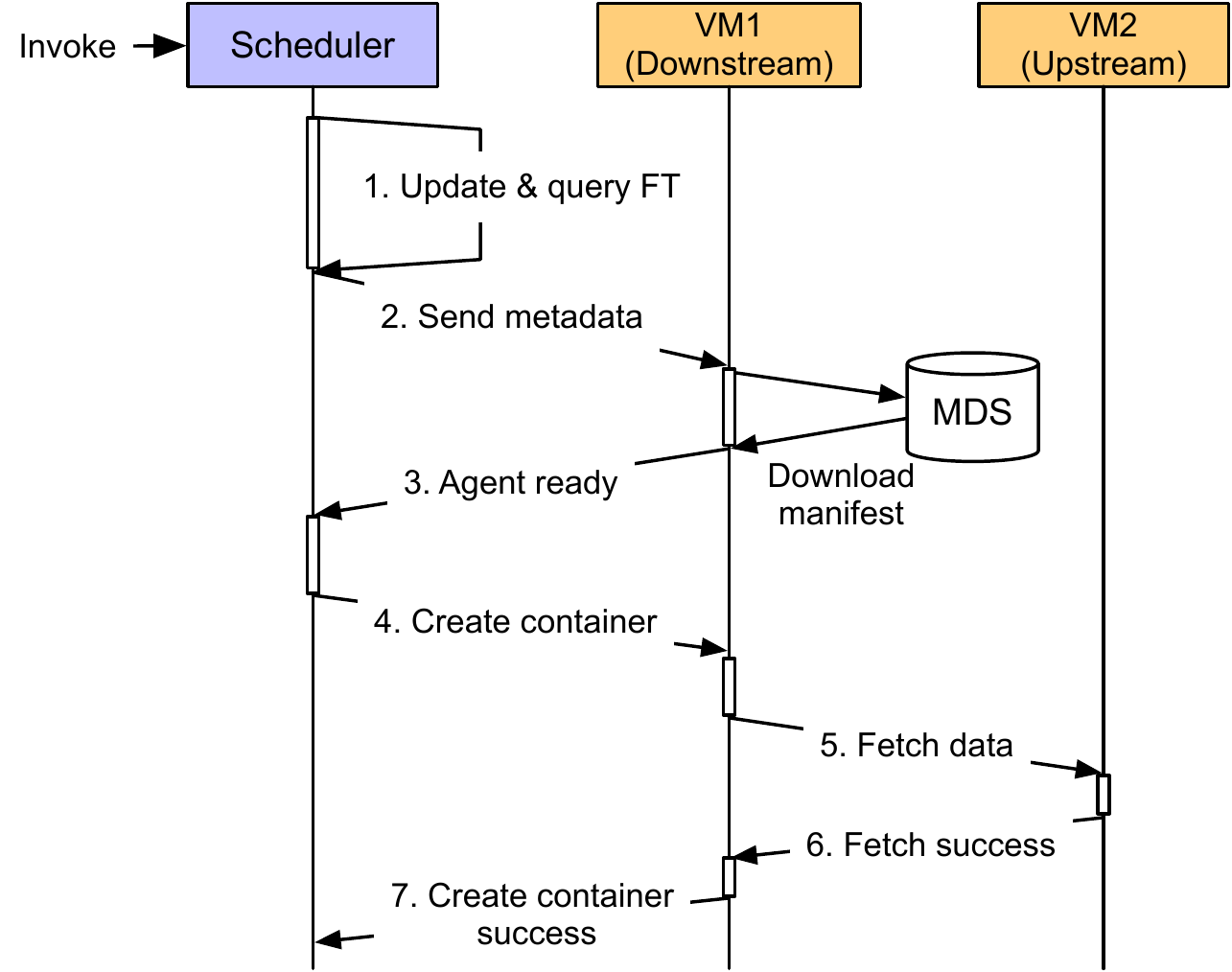}
\vspace{-10pt}
\caption{Container provisioning protocol. 
MDS: metadata store.
}
\label{fig:config_route}
\end{center}
\vspace{-20pt}
\end{figure}

\vspace{-12pt}
\paragraph{Container Provisioning Protocol.}
We design a protocol to coordinate the RPC communications between the scheduler and {\proj} VM workers and facilitate container provisioning (Figure~\ref{fig:config_route}). On an invocation request, if the scheduler detects that there are not enough active VMs to serve the request or all of current VMs are busy serving requests, the scheduler reserves one or multiple new VMs from the free VM pool and then enter the container provisioning process. 
Without  loss  of  generality,
we assume only one VM ($VM_1$) is reserved in this case.
In Step 1, the scheduler creates a new metadata object for $VM_1$ and inserts it to the FT associated with the requested \textit{$function_{ID}$}. The scheduler then queries the FT in order to get the \textit{address:port} of the upstream peer VM ($VM_2$).  
In Step 2, the scheduler sends the function metadata  
and \textit{address:port} of $VM_2$ to $VM_1$. Once receiving the information, $VM_1$ performs two tasks: 
(1)~downloads the {\small\texttt{.tar}} manifest file of the function container image from the metadata store (\cref{subsec:overview}),
and (2)~loads and inspects the manifest, fetches the URLs of the image layers, and persists the URL information on $VM_1$'s local storage.
In Step 3, $VM_1$ replies back to the scheduler that it is ready to start creating the container runtime for the requested function.
The scheduler receives the reply from $VM_1$ and then sends a \textit{create container} RPC request to $VM_1$ in Step 4. 
In Step 5 and 6, $VM_1$ fetches the layers from upstream $VM_2$ based on the manifest configuration processed in Step 2. 
In Step 7, $VM_1$ sends the scheduler an RPC that the container has been created successfully.

\vspace{-12pt}
\paragraph{FT Fault Tolerance.}
The scheduler pings VMs periodically and can quickly detect VM failures. If a VM is down, the scheduler notifies the FT manager to perform tree balancing operations in order to fix the FT topology.



\vspace{-10pt}
\subsection{FT Design Discussion}
\label{subsec:ft_discussion}
\vspace{-6pt}

{\proj} offloads the metadata-heavy management tasks to the existing FaaS scheduler, 
so that each individual node in a FT serves
the same role of fetching data from its parent peer (and seeding data for its child nodes if any). FT's root node does not have a parent peer but instead fetches data from the registry. {\proj}'s FT design can completely eliminate the I/O traffic to the registry, as long as a FT has at least one active VM that stores the requested container. Earlier, our workload analysis reveals that a typical FaaS application would always have a throughput above 0 RPS (\cref{subsec:workload}). This implies that, in practice, it is more likely for a request burst to scale out a FT from 1 to $N$ rather than from 0 to $N$.

\begin{figure}
\begin{center}
\includegraphics[width=.48\textwidth]{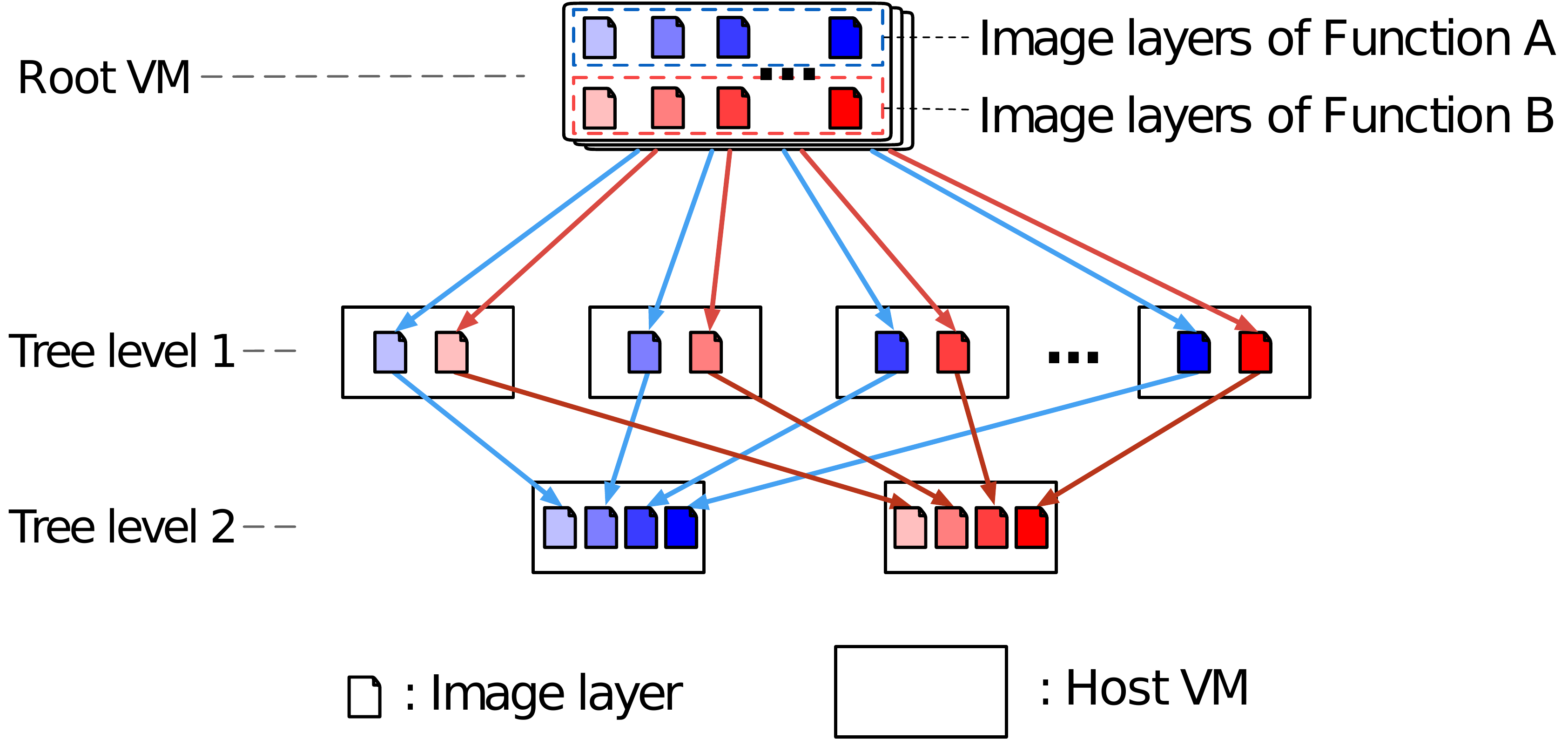}
\vspace{-15pt}
\caption{
An example tree that manages the topology at layer granularity and relies on root node for data seeding and tree management.
}
\label{fig:bp2p}
\end{center}
\vspace{-25pt}
\end{figure}

An alternative design is
to manage the topology at finer-grained layer (i.e., blobs) granularity. 
In this approach, each individual layer forms a logical layer tree; layers that belong to a function container image may end up residing on different VMs. Note that {\proj}'s FT is a special case of a layer tree model. 
Figure~\ref{fig:bp2p} shows an example. In this example, one VM stores layer files that belong to different function container images. Thus, a network bottleneck may occur when many downstream VM peers are concurrently fetching layers from this VM.
This is because many overlapping layer trees form a fully-connected, all-to-all network topology. An all-to-all topology might scale well if VMs are connected with high-bandwidth network. However, the all-to-all topology can easily create network bottlenecks if each VM is resource-constrained, which is our case in {\cloud}. We use small VMs with 2-core CPU, 4 GB memory, and 1 Gbps network in our FaaS infrastructure. 

Existing container distribution techniques~\cite{uber_kraken, DADI} rely on powerful root node to serve a series of tasks including data seeding, metadata management, and P2P topology management. Porting these frameworks to our FaaS platform would require extra, dedicated, possibly sharded, root nodes, which would add unnecessary cost to the provider. {\proj}'s FT design, on the other hand, keeps each VM worker's logic simple while offloading all logistics functions to our existing scheduler. This design naturally eliminates both the network I/O bottleneck and the 
root node bottleneck. 
In \cref{subsec:scalability} and \cref{subsec:neighbor} we evaluate and compare {\proj}'s FT design against a Kraken-like approach~\cite{uber_kraken, kraken_infoq}, which adopts a layer-based topology with powerful root nodes.

\vspace{-10pt}
\subsection{Optimizations}
\label{subsec:on_demand}
\vspace{-6pt}

We present the low-level optimizations that {\proj} uses to improve the efficiency of function container provisioning. 

\vspace{-12pt}
\paragraph{I/O Efficient Data Format.}
Regular {\small\texttt{docker pull}} and {\small\texttt{docker start}} are inefficient and time-consuming as the whole container image and the data of all the layers must be downloaded from a remote container registry~\cite{harter2016slacker} before the container can be started. To solve the issue,  we design a new block-based image fetching mechanism within {\cloud}. This mechanism uses an I/O efficient compression data file format.
Original data is split into fixed-sized blocks and compressed separately.
An offset table is used to record the offset of each compressed block in the compressed file. 


{\proj} uses the same data format for managing and provisioning code packages.
A code package is compressed into a binary file, which will be extracted by VM agent and eventually mounted inside of a function container. {\proj} distributes code packages the same way as it does for container images. \cref{subsec:fastlsmd} evaluates the performance benefit of I/O efficient data format on code package provisioning.

\vspace{-12pt}
\paragraph{On-Demand I/O.}
For applications that do not need to read all the layers at once on startup, our block-based image fetching mechanism provides them with an option to fetch layer data at fine-grained block level, in a lazy manner (i.e., on-demand), from a remote storage (in our case, a container registry or a peer VM). 
First, the application, in our case, a {\proj} VM worker, downloads the image manifest file from a metadata store and does an image load locally to load the {\small\texttt{.tar}} image manifest.
Second, it calculates the indices of the first and last (compressed) block and then consults with the offset table to find the offset information. Finally, it reads the compressed blocks and decompresses them until the amount of data that has been read matches the requested length. Since a read to the underlying (remote) block storage device must be aligned to the block boundary, the application may read and decompress more data than requested, causing read amplification.
However, in practice, decompression algorithm achieves much higher data throughput than that of a block storage or network. Thus, trading extra CPU 
overhead for reduced I/O cost is beneficial in our usage scenario. We evaluate the effectiveness of on-demand I/O in \cref{subsec:downloads}.

\vspace{-12pt}
\paragraph{RPC and Data Streaming.}
We build a user-level, zero-copy RPC library. This approach leverages non-blocking TCP {\small\texttt{sendmsg}} and {\small\texttt{recvmsg}} for transferring an {\small\texttt{struct iovec}} incontinuous buffer. The RPC library adds an RPC header directly to the buffer to achieve efficient, zero-copy serialization in the user space. 
The RPC library tags requests in order to achieve request pipelining and out-of-order receiving, similar to HTTP/2's multiplexing~\cite{http2}.
When a {\proj} worker receives a data block in its entirety, the worker immediately transfers the block to the downstream peer.

\vspace{-12pt}
\section{Evaluation}
\label{sec:eval}
\vspace{-6pt}

\begin{figure*}[t]
\begin{minipage}{\textwidth}

\begin{minipage}[b]{0.545\textwidth}
\begin{center}
\includegraphics[width=1\textwidth]{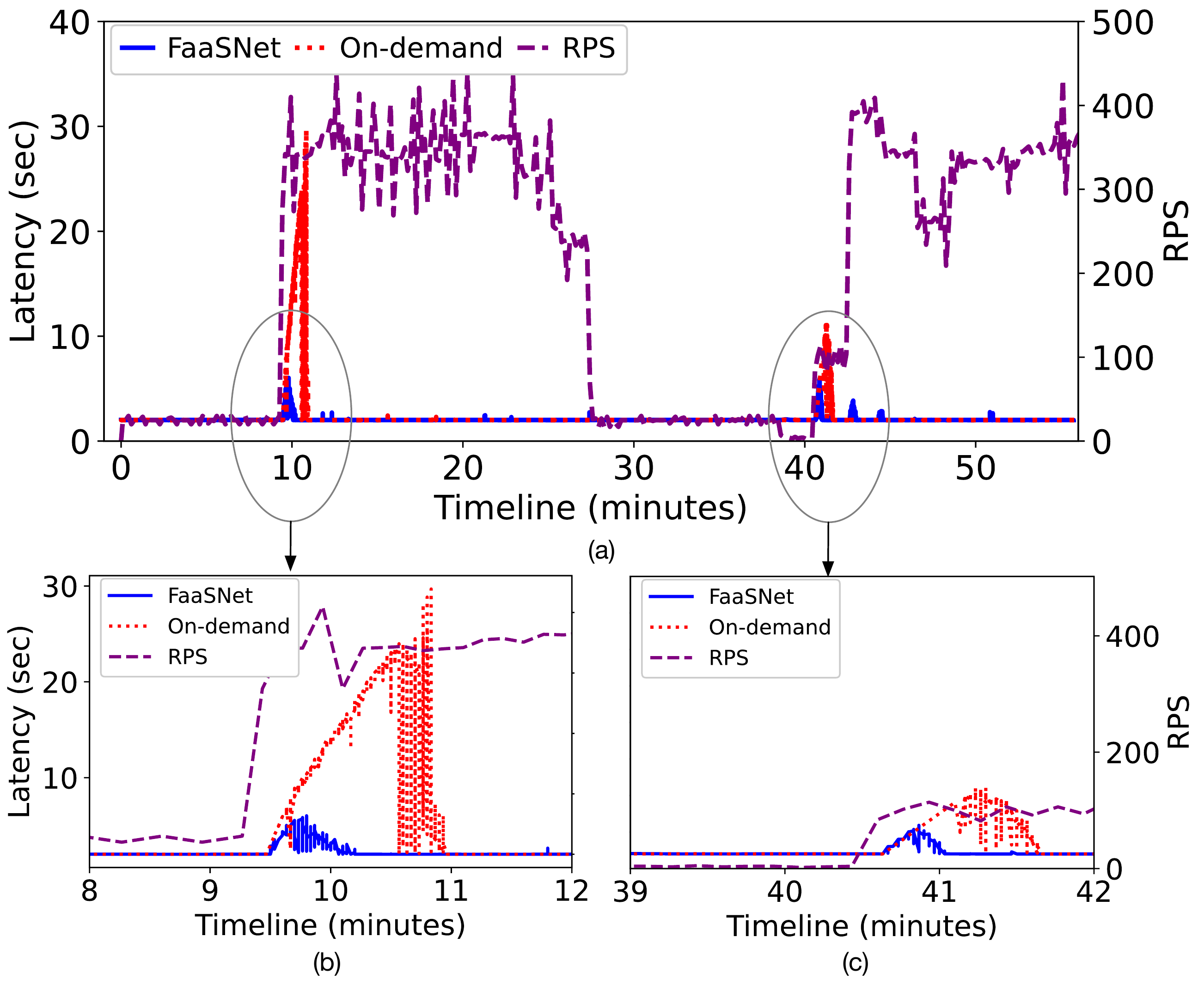}
\end{center}
\vspace{-20pt}
\caption{
IoT trace timeline.
}
\label{fig:iot-trace}
\end{minipage}
\hfill
\begin{minipage}[b]{0.438\textwidth}
\begin{center}
\includegraphics[width=1\textwidth]{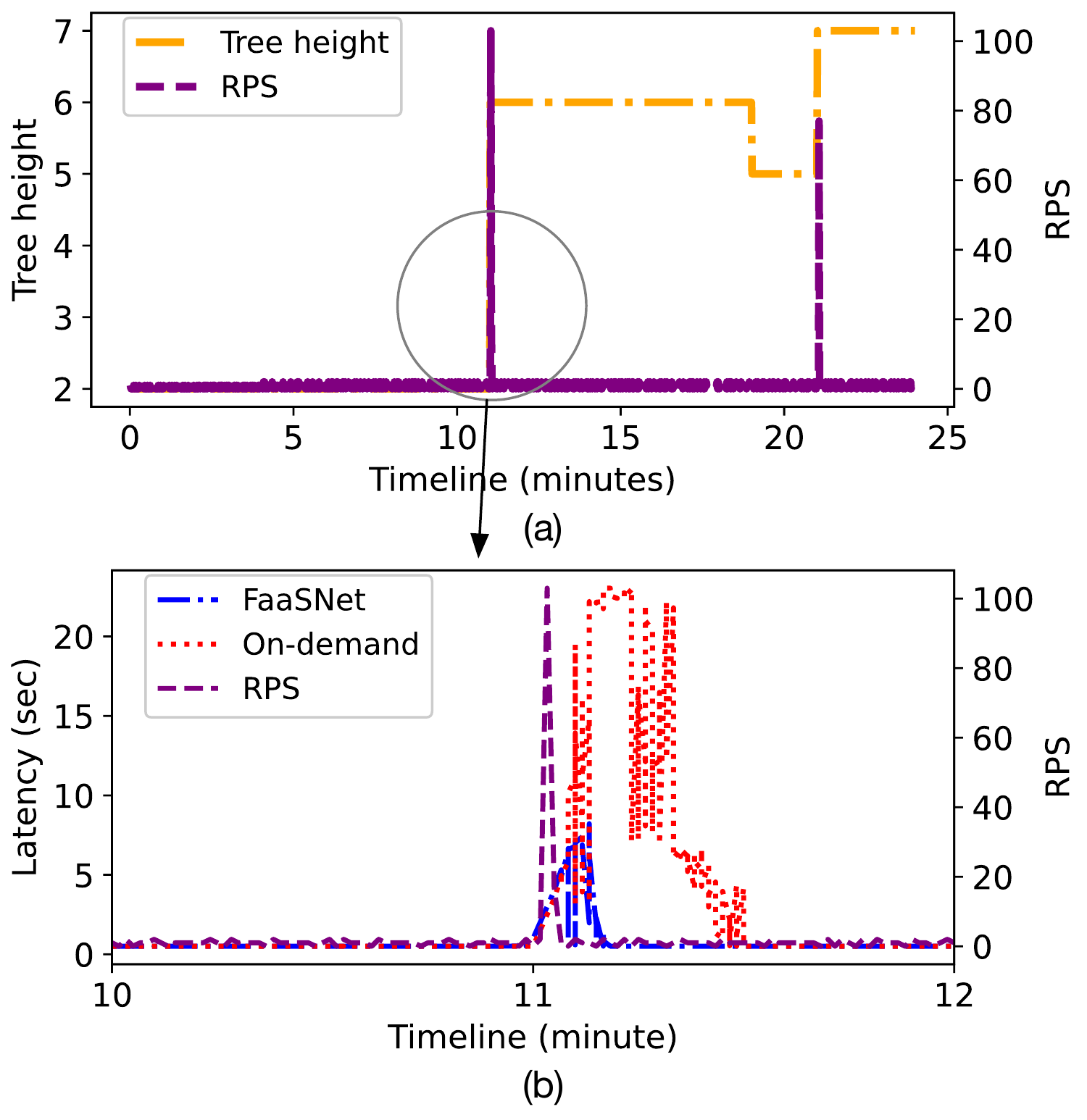}
\end{center}
\vspace{-20pt}
\caption{
Synthetic trace timeline. 
}
\label{fig:gaming}
\end{minipage}

\end{minipage}
\vspace{-20pt}
\end{figure*}

In this section, we evaluate {\proj} using production traces from {\cloud}'s FaaS platform. We also validate {\proj}'s scalability and efficiency via microbenchmarks. 



\vspace{-12pt}
\subsection{Experimental Methodology}
\vspace{-6pt}

We deploy {\proj} in {\cloud}'s {\fc} platform using a medium-scale, 500-VM pool and a large-scale, 1,000-VM pool. 
We follow the same deployment configurations used by our production FaaS platform:
all VMs use an instance type with 2 CPUs, 4 GB memory, 1 Gbps network;
we maintain a free VM pool where {\proj} can reserve VM instances to launch cloud functions. 
This way, the container provisioning latency does not include the time to cold start a VM instance. {\proj} uses a block size of 512 KB for on-demand  fetching and streaming.
Unless otherwise specified: we use a function that runs a Python 3.8 PyStan application for about 2 seconds; the size of the function container image is 758 MB; the function is configured with 3008 MB memory; each VM runs one containerized function.

\vspace{-12pt}
\paragraph{System Comparison.}
In our evaluation, we compare {\proj} against the following three configurations:
\vspace{-4pt}
\begin{enumerate}
    \item {\bf {\kraken}:} Uber's P2P-based registry system~\cite{uber_kraken, kraken_infoq}.
    We deploy a {\kraken} {\small\texttt{devcluster}}~\cite{kraken_dev_github}  with one origin node on our resource-constrained VM infrastructure.
    \vspace{-6pt}
    \item {\bf {\base}:} {\cloud} {\fc}'s current production setup. {\base} downloads container images using vanilla {\small\texttt{docker pull}} from a centralized container registry.
    \vspace{-6pt}
    \item {\bf {\opt}:} An optimized system based on \emph{baseline} but fetches container layer data on demand (\cref{subsec:on_demand}) from the container registry.
    \vspace{-6pt}
    \item {\bf{\optt}:} Alibaba's DADI~\cite{DADI, dadi_github} with P2P enabled. This approach uses one resource-constrained VM as the root node to manage the P2P topology.
    \vspace{-6pt}
\end{enumerate}
\vspace{-6pt}

\vspace{-6pt}
\paragraph{Goals.}
We aim to answer the following questions:
\vspace{-4pt}
\begin{enumerate}
    \item Can {\proj} rapidly provision function containers under bursty FaaS workloads with minimum impact on workload performance (\cref{subsec:eval_trace})?
    \vspace{-6pt}
    \item Does {\proj} scale with increasing invocation concurrency levels (\cref{subsec:scalability})?
    \vspace{-6pt}
    \item How does function placement impact {\proj}'s efficiency (\cref{subsec:neighbor})?
    \vspace{-6pt}
    \item 
    How does {\proj}'s 
    I/O efficient data format perform (\cref{subsec:fastlsmd})?
    \vspace{-6pt}
    \item How effective is {\proj}'s on-demand fetching (\cref{subsec:downloads})?
\end{enumerate}
\vspace{-6pt}

\vspace{-8pt}
\subsection{FaaS Application Workloads}
\label{subsec:eval_trace}
\vspace{-6pt}

In this section, we evaluate {\proj} using (scaled-down) application traces collected from our production workload (detailed in \cref{subsec:workload}). 

\begin{figure}
\begin{center}
\subfigure[IoT trace.] {
\includegraphics[width=.23\textwidth]{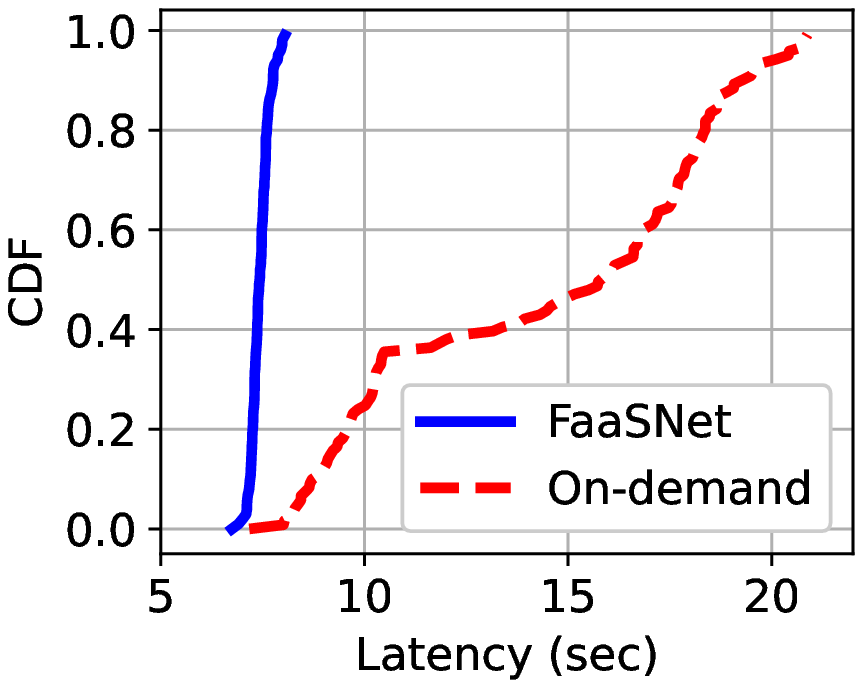}
\label{fig:sample_trace_container}
}
\hspace{-12pt}
\subfigure[Synthetic trace.] {
\includegraphics[width=.23\textwidth]{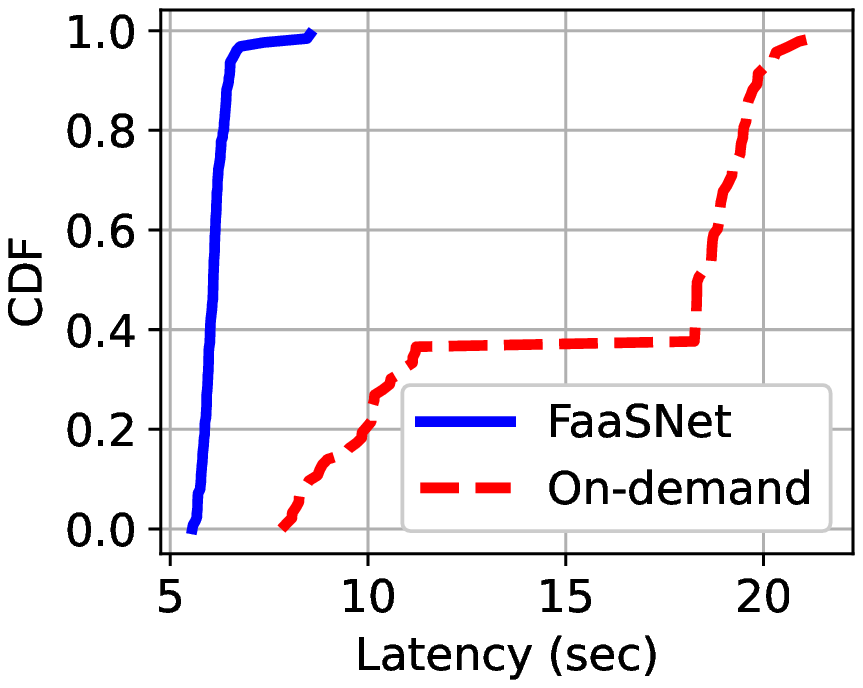}
\label{fig:iot}
}
\vspace{-12pt}
\caption{Distribution of container provisioning latency.}
\label{fig:trace_create_container}
\end{center}
\vspace{-20pt}
\end{figure}

\vspace{-12pt}
\paragraph{Trace Processing and Setup.}
We evaluate {\proj} using two FaaS applications: an IoT app and a gaming app. 
Since the original gaming workload exhibits a gradual ramp-up in throughput (Figure~\ref{fig:thpt_trace}(a)), we instead create a synthetic bursty workload based on the gaming workload  to simulate a sharp burst pattern for stress testing purpose. Our testing cluster has up to 1,000 VMs, so we scale down the peak throughput of both workload traces proportional (about $1/3$ of the original throughput) to our cluster size and shorten the duration from 2 hours to less than 1 hour. 

\vspace{-12pt}
\paragraph{IoT Trace.}
The IoT trace exhibits two invocation request bursts. The first burst happens at 9 minute and the throughput increases from 10 RPS to 300-400 RPS; the peak throughput lasts for about 18 minutes and returns back to 10 RPS at 28 minute. 
The second burst happens at 40 minute and the throughput increases to 100 RPS, and then in about 2 minutes, jumps to around 400 RPS. Figure~\ref{fig:iot-trace}(a) plots the 55-minute timeline of the workload's throughput and latency changes. 

\begin{figure*}[t]
\begin{center}
\subfigure[Average function container provisioning latency.
] {
\includegraphics[width=.48\textwidth]{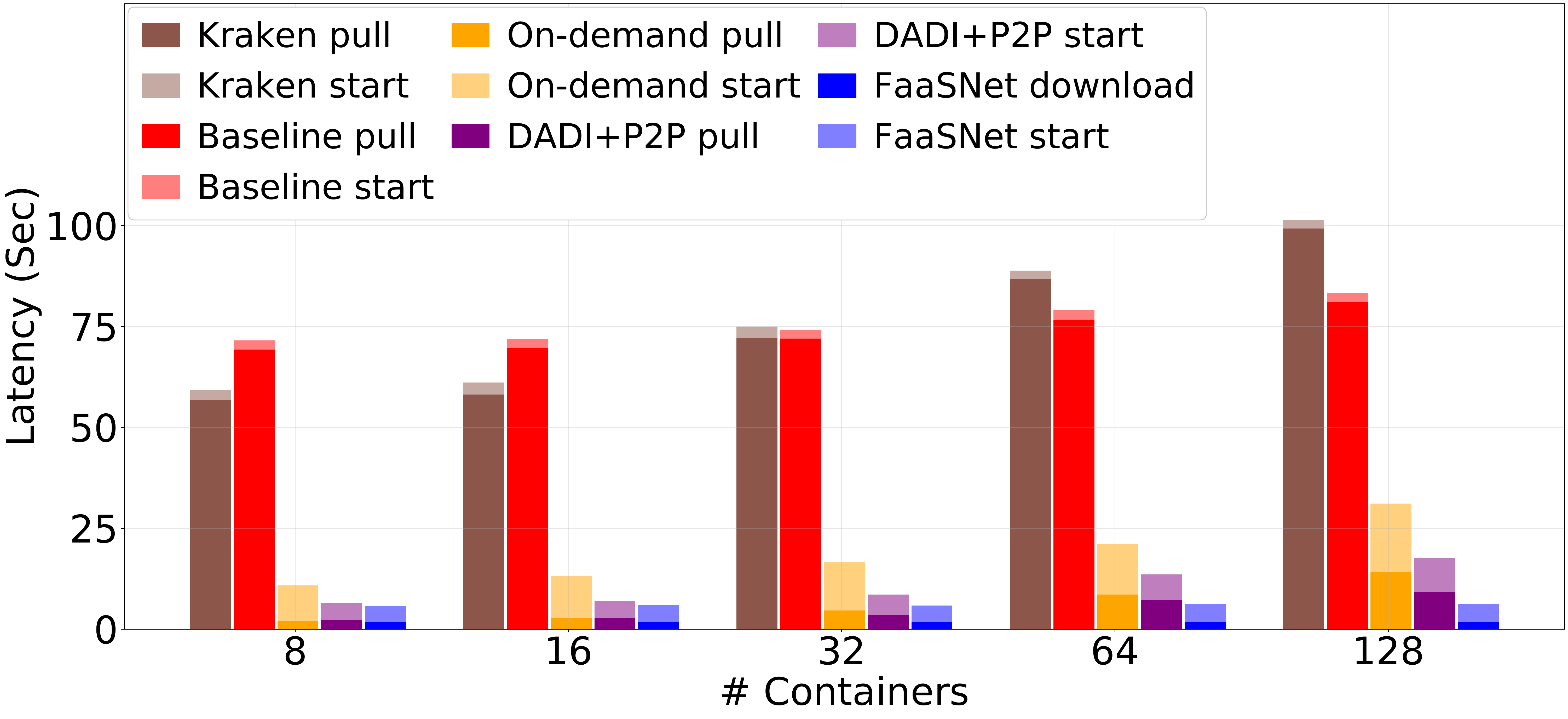}
\label{fig:scalability}
}
\hspace{-5pt}
\subfigure[Fraction of time spent at different stages.] {
\includegraphics[width=.48\textwidth]{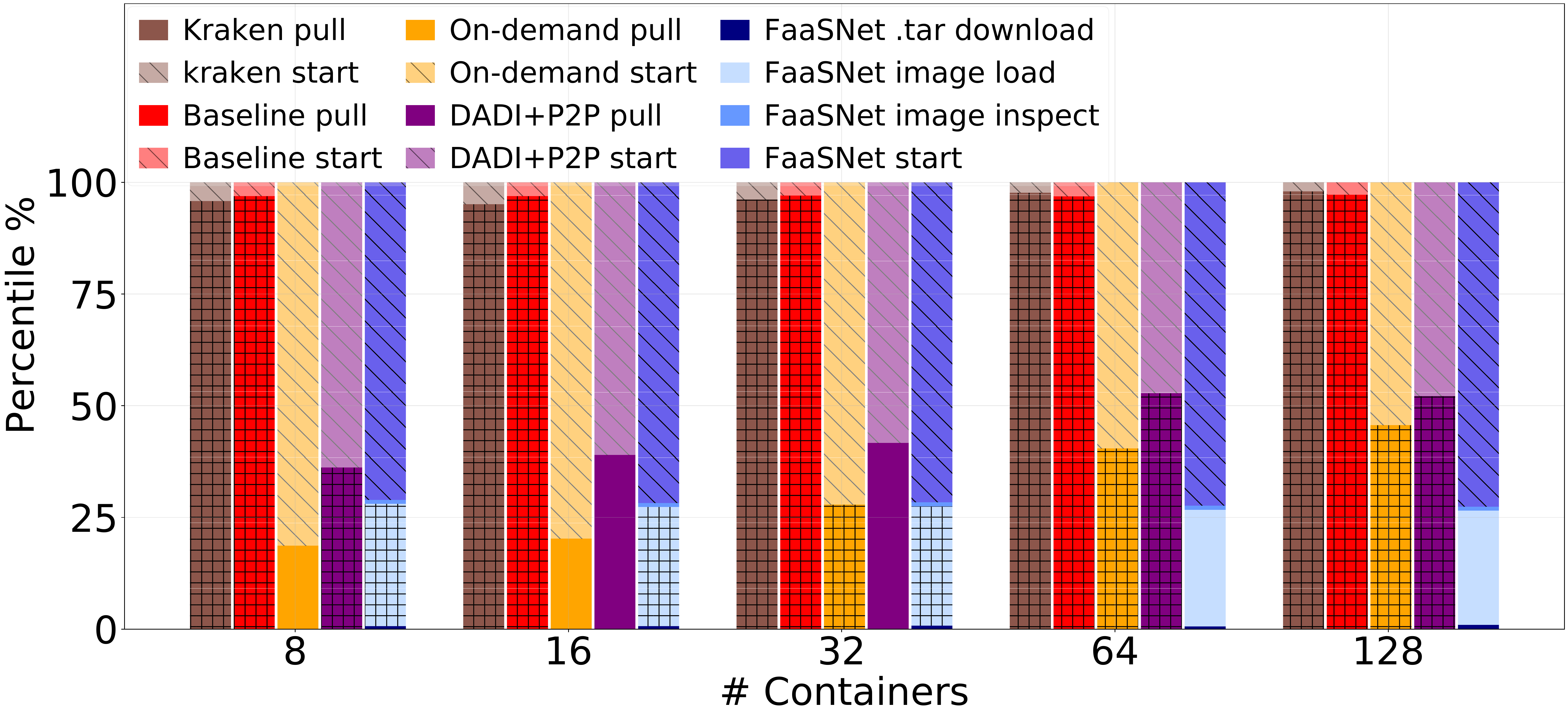}
\label{fig:breakdown}
}
\vspace{-10pt}
\caption{Container provisioning scalability test. 
}
\label{fig:micro}
\end{center}
\vspace{-20pt}
\end{figure*}

At 10 minute, the instantaneous throughput increase causes a backlog of function invocation requests at the FaaS scheduler side. Thus, the scheduler scales out the active VM pool 
by reserving a large number of free VMs from the free VM pool and starts the function container provisioning process. In {\base} case, all newly reserved VMs start pulling container images from the registry, which creates a performance bottleneck 
at the registry side. As a result, the application-perceived response time---the end-to-end runtime that includes the container startup latency and the function execution time of around 2 seconds---increases from 2 seconds to about 28 seconds. Worse, the registry bottleneck inevitably prolongs the time that {\base} requires to bring the response time back to normal. As shown in Figure~\ref{fig:iot-trace}(b), {\base} 
finishes the whole container provisioning process and brings the response time back to normal in almost $113$ seconds. 

In contrast, {\proj} avoids the registry bottleneck---instead 
of downloading the container image from the registry, each newly reserved VM fetches image data block-by-block from its upstream peer in the FT, forming a data streaming pipeline. 
As long as a VM fetches enough data blocks, it starts the container. 
{\proj} reduces the maximum response time from 28 seconds to 6 seconds. Out of the 6 seconds, around 4 seconds are spent on fetching image layers from the upstream peer VM. (We present the  container provisioning latency later in Figure~\ref{fig:trace_create_container}.)
More importantly,  {\proj} requires only $28$ seconds to bring the service back to normal, an improvement of $4\times$ compared to the {\opt} case.

\vspace{-12pt}
\paragraph{Synthetic Trace.}
In the synthetic trace test, we simulate two function invocation request bursts and evaluate FT's adaptivity. Figure~\ref{fig:gaming}(a) shows the timeline of a {\proj} FT's height changes. 
At 11 minute, the throughput suddenly grows from 1 RPS to 100 RPS. {\proj} detects the burst and rapidly scales the FT from a height of 2 (one root VM and one peer VM) to 7 (82 VMs in total). The FT starts parallel container provisioning instantly at 11 minute and sustains the latency spikes in about 10 seconds (Figure~\ref{fig:gaming}(b)). 
After the first burst, the throughput drops back to 1 RPS. Some VMs become cold and get reclaimed by the VM manager in about 15 minutes since the first burst. 
The number of VMs gradually reduces to 30 before the second burst arrives. Correspondingly, the height of the FT reduces from 6 to 5 (Figure~\ref{fig:gaming}(a)). When the second burst comes at 21 minute, the FT manager decides to grow the FT by adding another 62 VMs. With a total of 102 VMs, the height of the FT reaches up to 7 for serving the concurrent requests of the second burst.

\vspace{-12pt}
\paragraph{Container Provisioning Cost}
We next analyze the container provisioning latency seen 
in the two workloads. As shown in Figure~\ref{fig:trace_create_container}, since the registry in {\opt} incurs a performance bottleneck, {\opt} sees highly variant container provisioning latency, ranging from around 7 seconds to as high as 21 seconds. About $80\%$ of the containers take at least 10 seconds to start. 
The container startup latency is highly predictable in {\proj}, with significantly less variation.
For the synthetic workload, around $96\%$ of the functions require only 5.8 seconds to start. For the IoT workload, almost all the functions start execution within a short time range between 6.8-7.9 seconds. 
This demonstrates that {\proj} can achieve predictable container startup latency.

\vspace{-12pt}
\subsection{Scalability and Efficiency}
\label{subsec:scalability}
\vspace{-6pt}

Next, we evaluate {\proj}'s scalability and efficiency via microbenchmarking. 

\vspace{-12pt}
\paragraph{Scaling Function Container Provisioning.}
In this test, we measure the time {\proj} takes to scale from $0$ to $N$ concurrent invocation requests, where $N$ ranges from 8 to 128.
Each invocation request creates a single container in a VM. Figure~\ref{fig:micro} reports the detailed results. 
As shown in Figure~\ref{fig:scalability}, 
{\kraken} performs slightly better than {\base} under 8 and 16 concurrent requests but scales poorly under 32-128 concurrent requests. This is because 
{\kraken} distributes containers at layer granularity using a complex, all-to-all, P2P topology, which creates bottlenecks in the VMs. 
{\kraken} takes 100.4 seconds to launch 128 containers.

{\base} achieves slightly better scalability than {\kraken}.
The average container provisioning latency reaches up to 83.3 seconds when {\base} concurrently starts 128 functions. 
Two factors contribute to the delay: 
(1)~the registry becomes the bottleneck, 
and (2)~{\base}'s {\small\texttt{docker pull}} must pull the whole container image and layers (758 MB worth of data) from the registry and extract them locally. 

Adding {\opt} container provisioning to {\base} improves the latency significantly. This is because {\opt} eliminates most of the network I/Os for image layers that will not be instantly needed container startup. Despite pulling much less amounts of data from the registry, {\opt} still suffers from the registry bottleneck; provisioning 128 function containers requires $2.9\times$ longer time than provisioning 8 containers in {\opt} system.

{\optt} enables VMs to directly fetch image layers from peers, further avoiding downloading a large amount of layer blocks from the registry.
However, {\optt} still has two bottlenecks: one at the registry side---image pulls are throttled at the registry, and layer-wise extract operation may also be delayed in a cascading manner by local VMs;
the other bottleneck at the P2P root VM side---in addition to seeding data, the root VM in {\optt} is responsible for a series of extra tasks such as layer-tree topology establishment and coordination, thus forming a performance bottleneck.
This can be evidenced from Figure~\ref{fig:breakdown} that the fractions of {\optt}'s image pull and container start maintain at the same level when scaling from 64 to 128 function starts.

Figure~\ref{fig:scalability} shows that {\proj} scales perfectly well under high concurrency and achieves a speedup of $13.4\times$ than {\base} and $16.3\times$ than {\kraken}. {\proj} is $5\times$ and $2.8\times$ faster than {\opt} and {\optt}, respectively. 
As shown in Figure~\ref{fig:breakdown}, {\proj}'s average container provisioning latency is dominated by two operations: image load and container start.
{\proj} eliminates the bottleneck on both the two operations---on image load, {\proj} enables decentralized image loading (functionality-wise equivalent to image pull) at each VM by allowing each {\proj} worker to fetch the image manifest from the metadata store (with negligible overhead) and then starting the image loading process locally in parallel;
on  container start, each {\proj} VM worker directly fetches layer blocks from peer VM and starts the function container once enough blocks are fetched.
With all these optimizations, {\proj} maintains almost identical latency when scaling from 8 to 128 function startups.


\begin{figure}
\begin{center}
\includegraphics[width=.42\textwidth]{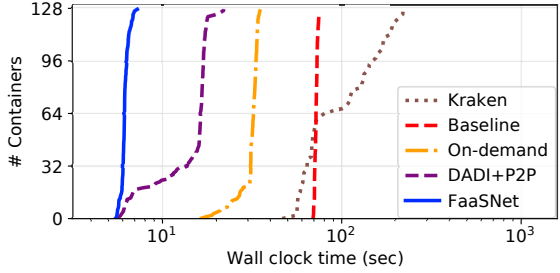}
\vspace{-10pt}
\caption{
Container provisioning scalability test: wall clock time (X-axis) for starting $N$ functions (Y-axis).
}
\label{fig:startup-duration}
\end{center}
\vspace{-15pt}
\end{figure}

\begin{figure}
\begin{center}
\includegraphics[width=.42\textwidth]{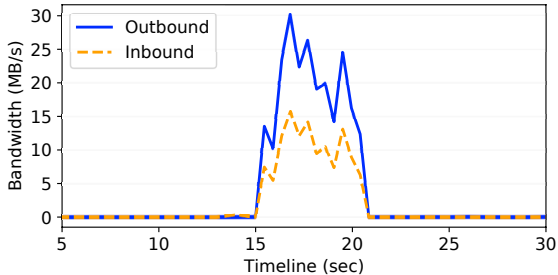}
\vspace{-10pt}
\caption{
A timeline of the VM network bandwidth usage. 
}
\label{fig:bandwidth}
\end{center}
\vspace{-20pt}
\end{figure}

\vspace{-12pt}
\paragraph{Function Container Provisioning Pipeline.}
We next examine how long the whole container provisioning process spans. Figure~\ref{fig:startup-duration} plots the timeline process that each system goes through to start $N$ function containers. 
We only report the 128-function concurrency case. 
We observe that {\proj} starts the first function at 5.5 second and the $128^{th}$ function at 7 second respectively. The whole container provisioning process spans a total of 1.5 seconds. Whereas {\opt} and {\optt} span a total duration of $16.4$ and $19$ seconds, respectively.
Specifically, it takes {\optt} a total of 22.3 seconds to start all the 128 containers, which is $14.7\times$ slower than that of {\proj}. This demonstrates that {\proj}'s FT-based container provisioning pipeline incurs minimum overhead and can efficiently bring up a large amount of function containers almost at the same time.

Figure~\ref{fig:bandwidth} shows the bandwidth usage timeline for a VM that we randomly select from the 128-function concurrency test. Recall that a {\proj} worker along a FT path (i.e., not the root VM nor the leaf VM) performs two tasks: (1) fetches layer data from the upstream VM peer, and (2) seeds layer data to the two children VM peers in its downstream paths. We observe that the bandwidth usage of the inbound connection (fetching layers from upstream) is roughly half of that of the two outbound connections (sending layers to downstreams) during container provisioning. The aggregate peak network bandwidth is 45 MB/s, which is $35.2\%$ of the maximum network bandwidth of the VM. We also observe that, the outbound network transfer is almost perfectly aligned with the inbound network transfer, again demonstrating the efficacy of {\proj}'s block-level data streaming scheme.

\begin{figure}[t]
\begin{center}
\includegraphics[width=.42\textwidth]{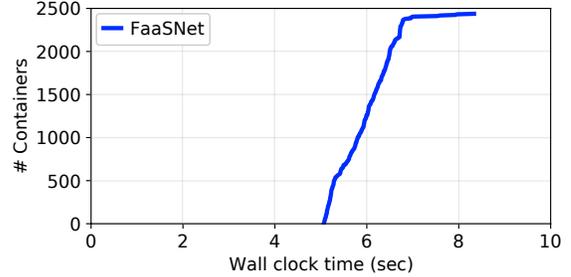}
\vspace{-10pt}
\caption{
Large-scale function container provisioning:
the wall clock time (X-axis) for starting $N$ functions (Y-axis).
}
\label{fig:large-scale}
\end{center}
\vspace{-15pt}
\end{figure}

\vspace{-12pt}
\paragraph{Large-Scale Function Startup.}
In this test, we create $1,000$ VMs and concurrently invoke $2,500$ functions on them. Each function uses a container of 428 MB and is configured to run with 1024 MB memory. 
Each VM runs two or three functions in this test.
Figure~\ref{fig:large-scale} shows that all function containers finish provisioning and start running between 5.1 second and 8.3 second,
again demonstrating {\proj}'s superb scalability. None of {\opt} and {\optt} finishes the test due to timeout errors.

\begin{figure}
\begin{center}
\includegraphics[width=.42\textwidth]{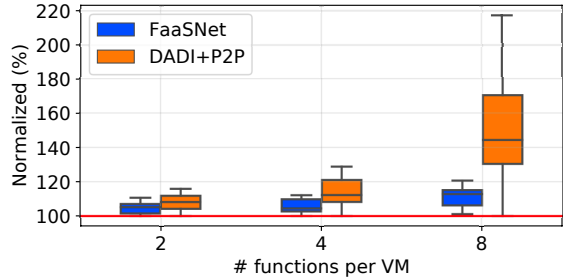}
\vspace{-10pt}
\caption{
Container provisioning latency as a function of various function placement situations.
For {\proj} and {\optt}, the latency is normalized to that of provisioning a single container in one VM in their own case.
}
\label{fig:neighbor}
\end{center}
\vspace{-20pt}
\end{figure}

\vspace{-14pt}
\subsection{Impact of Function Placement}
\label{subsec:neighbor}
\vspace{-4pt}

We conduct a sensitivity analysis to quantify the impact of function placement on container provisioning. In this test, we concurrently invoke 8 functions on $N$ VMs, where $N$ varies from 4 to 1. Each function has a different container (75.4 MB) and is configured to use 128 MB function memory (since a VM has 4 GB memory, it is allowed to host as much as 20 functions with 128 MB memory). We compare the container provisioning latency between {\proj} and {\optt}. As shown in Figure~\ref{fig:neighbor}, {\optt} sees much higher latency variation when 4 functions and 8 functions are placed on the same VM, because {\optt}'s root VM is overloaded by the establishment processes of many small layer trees. 


\begin{figure}[t]
\begin{center}
\subfigure[End-to-end latency.]{
\includegraphics[width=.22\textwidth]{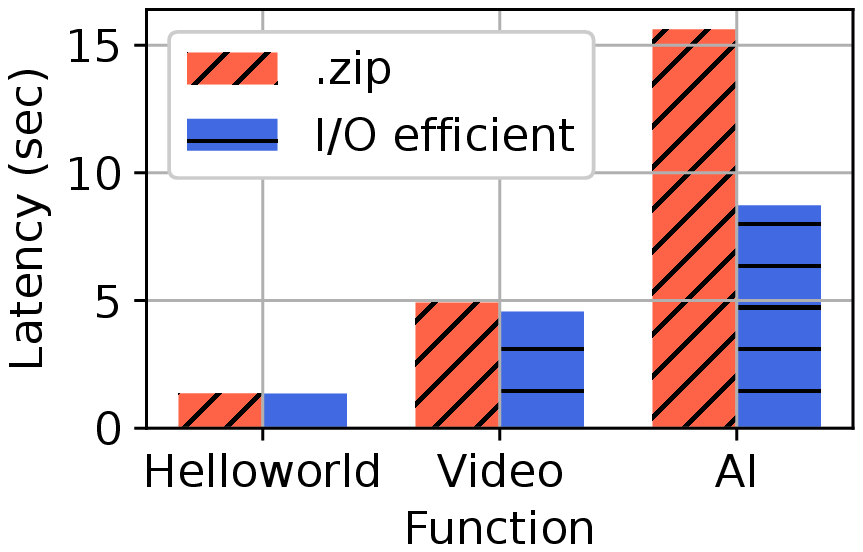}
\label{fig:code_package_latency}
}
\hspace{-8pt}
\subfigure[Code package size.] {
\includegraphics[width=.22\textwidth]{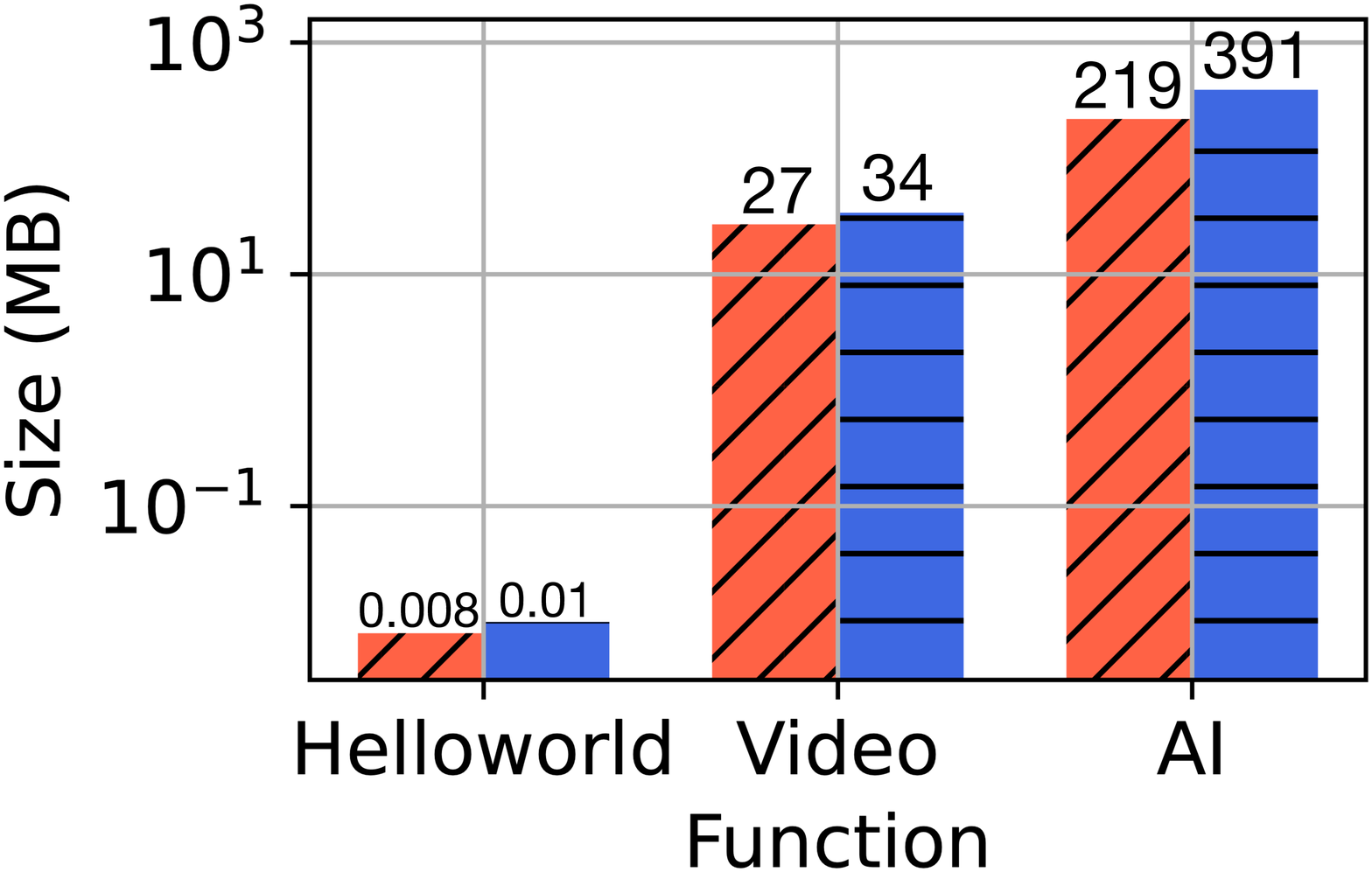}
\label{fig:code_package_size}
}
\vspace{-10pt}
\caption{End-to-end invocation latency and code package size comparison between {\fastlsmd} and \small\texttt{.zip}.}
\label{fig:code_package}
\end{center}
\vspace{-20pt}
\end{figure}

\vspace{-12pt}
\subsection{I/O Efficient Data Format}
\label{subsec:fastlsmd}
\vspace{-4pt}

We next evaluate how the I/O efficient format helps with code package provisioning.
We choose three functions: a simple, Python-written HelloWorld function that sleeps for 1 second (Helloworld), an FFmpeg video encoding function (Video), and a TensorFLow Serving function (AI), and compare {\proj}'s {\fastlsmd} with the baseline {\small\texttt{.zip}} format. 

Figure~\ref{fig:code_package_latency} plots the end-to-end function invocation performance  including the latency of code package downloading and function duration. Our {\fastlsmd} performs the same as {\small\texttt{.zip}} for Helloworld, since Helloworld's code package has only 11 KB in size (Figure~\ref{fig:code_package_size}). 
The {\fastlsmd} achieves better performance compared to {\small\texttt{.zip}} for Video and AI since the {\fastlsmd} fetches data on demand rather than extracting all data as {\small\texttt{.zip}} does. 
Figure~\ref{fig:code_package_size} shows the code package sizes. Functions have a larger code package size when using {\fastlsmd}, because {\fastlsmd}'s compression incurs extra storage overhead.



\vspace{-12pt}
\subsection{On-Demand I/O: Sensitivity Analysis}
\label{subsec:downloads}
\vspace{-4pt}

Finally, we evaluate on-demand I/O and compare the impact of block sizes on read amplification.
With on-demand fetching, a {\proj} worker only needs to fetch enough layer data blocks in order to start the function container. 
We choose three different function container images:
(a)~a 195 MB helloworld image with a Python 3.9 runtime pulled from Docker Hub;
(b)~a 428 MB PyStan image based on an AWS Lambda Python 3.8 base image; 
and (c)~a 728 MB PyStan image based on an {\cloud} Python 3.8 base image.

\begin{figure}
\begin{center}
\includegraphics[width=.42\textwidth]{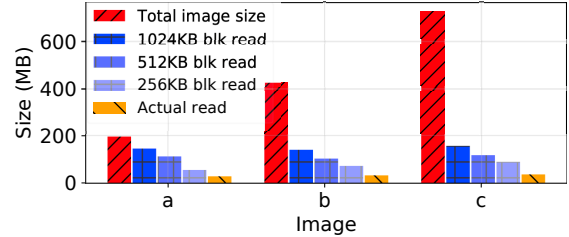}
\vspace{-10pt}
\caption{
Amounts of data fetched (on-demand) as a function of block sizes. 
Left-most bar in each bar cluster represents the size of the original container image;
right-most bar in each bar cluster represents the actual amount of data read from the data blocks fetched via network. 
}
\label{fig:block-size}
\end{center}
\vspace{-20pt}
\end{figure}

As shown in Figure~\ref{fig:block-size}, on-demand fetching can reduce the amount of data transferred via network. The reduction is especially profound for image b and c, because base images are dependency-heavy and are commonly used in the image building process.
For example, with a block size of 512 KB (the block size configuration that we use in our evaluation), on-demand fetching sees a $83.9\%$ reduction in network I/Os, compared to that of regular {\small\texttt{docker pull}}.

We also observe different levels of read amplification under different block sizes. 
This is because the starting and ending offset position is likely to be misaligned with the boundary of the (compressed) blocks in the underlying block device, the larger the block size is, the more useless data the {\proj} worker may read from the starting and ending blocks. 
The actual amount of data read (for starting a container) after decompression is even smaller, indicating that most of the dependencies included in the original container image is not used at the container startup phase. Exploring optimization to reduce the read amplification is part of our future work. 

\vspace{-12pt}
\section{Discussion}
\label{sec:discussion}
\vspace{-6pt}

In this section, we discuss the limitations and possible future directions of {\proj}.

\vspace{-12pt}
\paragraph{FT-aware Placement.} 
When the number of functions grows, the contention of network bandwidth ensues. Though \cref{subsec:neighbor} proves it less a concern in \proj{} than prior work, for the sake of safety in production, we program the system to avoid co-locating multiple functions if the cluster resources permit. Anticipating a future demand increase of custom containers, we plan to address the problem by extending the container placement logic. The general goal is to balance the inbound and outbound communication of each VM when multiple functions are being provisioned. Intuitively, by adjusting container placement, we can control the number of FTs that a VM is involved in and the role (e.g., leaf vs. interior node) a VM serves, and thus the bandwidth consumption. A further optimization is to co-locate functions that share common layers, so they could reduce the amount of data transfer.

\vspace{-12pt}
\paragraph{Multi-Tenancy.}
As mentioned, {\cloud} achieves strong, tenant-level function isolation using containers and VMs. As such, our FaaS platform cannot share VMs among tenants. This means that {\proj}'s FTs are naturally isolated between different tenants. Porting {\proj} to other secure and lightweight virtualization techniques~\cite{firecracker_nsdi20, lightvm_sosp17, nabla_socc18, xcontainer_asplos19} is our ongoing work.



\vspace{-12pt}
\paragraph{\proj{} for Data Sharing.} 
Technically, our work can enable the sharing of container images among VMs through P2P communication,
There is potentiality for it to generalize to a broader scope: data sharing for general container orchestration systems such as Kubernetes~\cite{k8s}. Such a need is arising in FaaS platforms with the emergence of data-intensive applications, such as matrix computation~\cite{numpywren_socc20, wukong_socc20}, data analytics~\cite{pywren_socc17, locus_nsdi19},  video processing~\cite{sprocket_socc18, excamera_nsdi17}, and machine learning~\cite{cirrus_socc19, lambdaml_sigmod21}, etc. Most of them rely on a centralized storage for data exchange, which is a similar bottleneck as the container registry in our work. Hence we believe the design of \proj{} can also accelerate data sharing, only with two additional challenges: (1) how to design a primitive interface for users; (2) how to adapt the tree management algorithms for more frequent topology building and change. We leave the exploration as a future work.


\vspace{-12pt}
\paragraph{Adversarial Workloads.}
Extremely short-lived functions with a duration at sub-second level and sparse invocations may be adversarial to {\proj} and custom-container-based FaaS platforms. Function environment caching and pre-provisioning~\cite{serverless_in_the_wild,sock_atc18, lambda_reuse} can be used to handle such workloads but with extra infrastructure-level costs.

\vspace{-12pt}
\paragraph{Portability.}
{\proj} is transparent to both upper-level FaaS applications and underlying FaaS infrastructure. It reuses {\fc}'s existing VM reclaiming policy and could be applied to other FaaS platforms without introducing extra system-level costs. Porting {\proj} to {\cloud}'s bare-metal infrastructure is our ongoing work.

\vspace{-10pt}
\section{Related Work}
\label{sec:related}


\vspace{-8pt}
\paragraph{Function Environment Caching and Pre-provisioning.}
FaaS applications face a notoriously persisting problem of high latency---the so-called ``cold start'' penalty---when function invocation requests must wait for the functions to start. Considerable prior work has examined ways
to mitigate the cold start latency in FaaS platforms. 
FaaS providers such as AWS Lambda and Google Cloud Functions pause and cache invoked functions for a fixed period of time to reduce the number of cold starts~\cite{peeking_atc18, lambda_reuse, infinicache_fast20}. This would, however, increase the TCO for providers.
To reduce such cost, researchers propose prediction methods that pre-warm functions just in time so that incoming recurring requests would likely hit on warm containers~\cite{serverless_in_the_wild}.
SAND shares container runtimes for some or all of the functions of a workflow for improved data locality and reduced function startup cost~\cite{sand_atc18}. 
SOCK caches Python containers with pre-imported packages and clones cached containers for minimizing function startup latency~\cite{sock_atc18}. 
PCPM pre-provisions networking resources and dynamically binds them to function containers to reduce the function startup cost~\cite{agile_cold_start_hotcloud19}.
While function requests can be quickly served using pre-provisioned, or cached, virtualized environments, these solutions cannot fundamentally solve the issue of high costs incurred during function environment provisioning. 

\vspace{-12pt}
\paragraph{Sandbox, OS, and Language-level Support.}
A line of work proposes low-level optimizations to mitigate FaaS cold start penalty.
Catalyzer~\cite{catalyzer_asplos20} and SEUSS~\cite{seuss_eurosys20} reduce the function initialization overhead by booting function instances from sandbox images created from checkpoints or snapshots. 
Systems such as Faasm~\cite{faasm_atc20} and \cite{micro_atc18} leverage lightweight language-based isolation to achieve speedy function startups. 
Unlike {\proj}, these solutions either require modified OSes~\cite{catalyzer_asplos20, seuss_eurosys20} or have limited compatibility and usability in terms of programming languages~\cite{faasm_atc20, micro_atc18}. 


\vspace{-12pt}
\paragraph{Container Storage.}
Researchers have looked at optimizing container image storage and retrieval. 
Slacker speeds up the container startup time by utilizing lazy cloning and lazy propagation~\cite{harter2016slacker}. Images are stored and fetched from a shared network file system (NFS) and referenced from a container registry. 
Wharf~\cite{wharf_socc18} and CFS~\cite{cfs_sigmod19} store container image layers in distributed file systems. Bolt provides registry-level caching for performance improvement~\cite{bolt_cloud19}. These work are orthogonal in that {\proj} can use them as backend container stores. 
Kraken~\cite{uber_kraken} and DADI~\cite{DADI} use P2P to accelerate container layer distribution. These systems assume a static P2P topology and require dedicated components for image storage, layer seeding, or metadata management, which leave them vulnerable to high dynamicity (demanding high adaptability of the network topology) and unpredictable bursts (requiring highly scalable container distribution). 

\vspace{-12pt}
\paragraph{AWS Lambda Containers.}
AWS announced the launch of container image support for AWS Lambda~\cite{aws_lambda_container} on December 01, 2020. Limited information was revealed via a re:Invent 2020 talk~\cite{aws_reinvent20} about this feature:
AWS uses multi-layer caching to aggressively cache image blocks: 
(1)~microVM-local cache, 
(2)~shared, bare-metal server cache, 
and (3)~shared, availability zone cache. 
The solution, while working for powerful, bare-metal server-based clusters that can co-locate many microVMs~\cite{firecracker_nsdi20}, is not suitable for our FaaS platform, which is based on thousands of small VMs managed by {\cloud}'s public cloud platform. 

\vspace{-12pt}
\paragraph{P2P Content Distribution.}
VMThunder uses a tree-structured P2P overlay for accelerating VM image distribution~\cite{vmthunder}. A BitTorrent-like P2P protocol is proposed for achieving similar goals~\cite{Chen_p2p}. Bullet uses an overlay mesh for high-bandwidth, cross-Internet file distribution~\cite{bullet_sosp03}. 
{\proj} builds on these works but differs with a new design that is attuned to the FaaS workloads. 

\vspace{-14pt}
\section{Conclusion}
\label{sec:conclusion}
\vspace{-6pt}

Scalable and fast container provisioning can enable fundamental elasticity for FaaS providers that support custom-container-based cloud functions. {\proj} is the first system that provides an end-to-end, integrated solution for FaaS-optimized container runtime provisioning. {\proj} uses lightweight, decentralized, and adaptive function trees to avoid major platform bottlenecks. {\proj} provides a concrete solution that is attuned to the requirements of a large cloud provider's ({\cloud}) FaaS platform. 
We show via experimental evaluation that {\proj} can start thousands of large function containers in seconds.
Our hope is that this work will make container-based FaaS platforms truly elastic and open doors to a broader class of dependency-heavy FaaS applications including machine learning and big data analytics.

To facilitate future research and engineering efforts, we release the source code of {\proj}'s FT prototype as well as an anonymized dataset containing production FaaS cold start traces
collected from {\cloud} {\fc} at: \url{https://github.com/mason-leap-lab/FaaSNet}.

\label{startofrefs}
\clearpage
\newpage
\balance

\vspace{-12pt}
\section*{Acknowledgments}
\vspace{-8pt}

We are grateful to our shepherd, Jon Crowcroft, and the anonymous reviewers, for their valuable feedback and suggestions. We would like to thank Benjamin Carver for his proofreading, and Jianlie Zou, Qi Sun, and Yifan Yuan for their kind help on this work. This work is sponsored in part by NSF under an NSF CAREER Award CNS-2045680. 

{
\bibliographystyle{plain}
\bibliography{refs}
}

\end{document}
